\documentclass{ametsocV6.1}
\nolinenumbers
\let\internallinenumbers\relax

\usepackage{graphicx}
\usepackage{amsmath}
\usepackage{amssymb}
\usepackage{subcaption}
\usepackage{float}

\title{GOES-East full-disk AI nowcasting of cloud evolution in observation space}

\authors{Dhamma Kimpara,\aff{a}\correspondingauthor{Dhamma Kimpara, dkimpara@ucar.edu}
Omid Bagheri,\aff{a}
Ivette Hernandez Banos,\aff{a}
Byoung-Joo Jung,\aff{a}
and Chris Snyder\aff{a}}

\affiliation{\aff{a}{NSF National Center for Atmospheric Research, Boulder, Colorado, USA}}
\abstract{Clouds affect aviation, solar energy, remote sensing, and storm prediction, yet they remain among the hardest atmospheric features to forecast, particularly at convective scales. Because clouds are shaped by processes spanning a wide range of space and time scales, numerical weather prediction, extrapolation methods, and existing machine learning (ML) approaches are each limited by some combination of accuracy, domain size, and temporal resolution. We present DOP+, an ML approach for clouds that forecasts GOES-East full-disk infrared brightness temperatures by extending direct observation prediction (DOP) with conditioning on meteorological fields. The domain covers tropical, midlatitude, and marine regimes across ${\sim}10^8$~km$^2$, roughly a fifth of Earth's surface. DOP+ forecasts cloud evolution at 10-minute resolution and outperforms persistence, synoptic-scale NWP, and a pure DOP baseline across 0--6~h lead times in fractions skill score and mean absolute error skill score. Convective structure is retained out to 2--3~h. DOP+ thus achieves a state-of-the-art combination of accuracy, temporal resolution, and spatial coverage. Our work lays the foundation for fully global cloud nowcasting at convective timescales.}

\begin{document}
\nolinenumbers

\maketitle
\nolinenumbers


\section{Introduction}


Cloud forecasting is of significance to aviation, solar energy, remote sensing, and storm prediction. However, accurately representing clouds in traditional physics-based Numerical Weather Prediction (NWP) models remains challenging. These challenges arise from limitations in the representation of convection, cloud microphysics parametrization, boundary layer parametrization, and data assimilation \citep{errico_assimilation_2007, sun_use_2014, morrison_confronting_2020}. While high-resolution, convective-scale NWP models can partly overcome these limitations, their computational cost makes them impractical for domains larger than mesoscale domains such as CONUS or Europe.

One could imagine that, with enough scientific and computational progress, we might eventually enable accurate global convective-scale NWP and, consequently, reliable cloud forecasting. In lieu of such a monumental and uncertain effort, we instead use machine learning to leverage large observational datasets directly.


We directly predict observations from GOES-East, a multi-spectral geostationary satellite, to capture cloud evolution \citep{schmit_chapter_2020}. These multi-spectral geostationary satellites capture snapshots of the atmosphere at high spatio-temporal resolution (2~km, ${\sim}$10~minute) where each channel is some function of the state of the atmosphere; e.g.\ vertically integrated moisture, cloud top temperature, and cloud composition. These satellite observations are used in operations in data assimilation, precipitation estimation, and by forecasters to analyze convection. Thus predicting these satellite observations offers a compelling way to forecast quantities of interest such as cloud cover, convective growth, and other diagnosable values.


Two challenges arise when predicting geostationary observations, and prior work has generally met one or the other but not both.

The first is temporal: clouds are governed by processes acting across a wide range of scales. Predicting clouds requires faithful representation of these multi-scale processes. In particular, fast processes, like convection, are among the least predictable and most consequential. Predicting them requires forecasting at a timescale of minutes rather than hourly or coarser; GOES-East provides a 10-minute cadence. The second is spatial scale; and it compounds upon the challenge of the first. A model over the GOES-East full disk must represent a large diversity of atmospheric regimes at once: tropical, mid-latitude, and marine regimes, and opposing seasonal regimes simultaneously, such as Northern-Hemisphere summer against Southern-Hemisphere winter. Such a domain presents a unique challenge that a limited-area model never faces.

Existing work addresses one challenge or the other. \citet{afzali_gorooh_deterministic_2026}, \citet{chase_score-based_2025}, \citet{chen_skillful_nodate}, \citet{dai_four-hour_2025}, \citet{kellerhals_nowcasting_2022}, and \citet{pathak_learning_2026} forecast geostationary observations over regional domains (CONUS or smaller) where a narrower range of regimes is in play. \citet{wei_dayu_2024} cover a global domain but do not resolve sub-hourly evolution. \citet{partio_cloudcasttotal_2025} targets total cloud cover derived from geostationary imagery, but is likewise regional and hourly. 


Our method draws mainly from direct observation prediction (DOP), an emerging paradigm in ML weather prediction (MLWP). While MLWP models trained and initialized from (re-)analyses have revolutionized weather forecasting, they are dependent on an existing data assimilation infrastructure for their training data and initialization \citep{alexe_graphdop_2024, pathak_learning_2026, pinnington_aifs-dop_2026}. The premise of DOP is that by initializing from and directly predicting observations, it avoids any dependence on the traditional data assimilation system. In fact, the hope is that this approach will improve, in a wholesale manner, on the limitations of the data assimilation system and the numerical model. Recent work has shown success in regional convective-scale and global synoptic medium-range forecasting (though it is overall unsettled how DOP compares to MLWP) \citep{alexe_graphdop_2024, chase_score-based_2025, chen_skillful_nodate, dai_four-hour_2025, kellerhals_nowcasting_2022, miralles_pointwise_2026, pathak_learning_2026, pinnington_aifs-dop_2026, wei_dayu_2024}.


For cloud prediction specifically, any emulator inherits the well-known deficiencies of NWP in representing clouds. DOP thus stands to make large gains, given the availability of high-resolution geostationary observations. Hence, many works have predicted geostationary observations directly targeting cloud-related features like storm and tropical cyclone prediction \citep{afzali_gorooh_deterministic_2026, chase_score-based_2025, chen_skillful_nodate, dai_four-hour_2025, kellerhals_nowcasting_2022, pathak_learning_2026, wei_dayu_2024}.


We augment the ``pure DOP'' approach with hourly meteorological input from a lower space- and time-resolution NWP product to inform the model of large-scale atmospheric conditions. We call this method DOP+ in this work. This is well-motivated due to the availability and accuracy of synoptic-scale forecasting products for synoptic-scale short range prediction. Such an approach has been explored by only two works to our knowledge \citep{chen_skillful_nodate, miralles_pointwise_2026}.

This type of hierarchical modeling is already the dominant approach in NWP practice: high-resolution regional models are forced on the boundary by coarse-resolution global models. Our approach also relies on a coarse forecast ``boundary'' input throughout the whole field. It takes it a step further by using the capabilities of ML to directly forecast the high-resolution observations.


DOP+ has analogues in other data-driven approaches to weather prediction. Like extrapolation approaches\footnote{These techniques are often termed ``nowcasting.'' Since that term is also used to describe 0--6~h forecasts by any method \citep{sun_use_2014, wmo_guidelines_2017}, we avoid its use here.} such as optical flow \citep[see][and references therein]{king_optical_2026}, DOP+ utilizes the latest observations. DOP+ can also be viewed as a way to augment an existing global weather prediction (whether from a NWP or MLWP) with detailed cloud information and thus resembles a prognostic version of traditional post-processing methods such as those that have shown success in predicting instantaneous convective hazards \citep{hua_improving_2025, sobash_evaluating_2024}. Our method therefore occupies the space between auto-regressive DOP and post-processing, a combination that has barely been explored in MLWP outside of large, expensive, and data-hungry foundation model approaches.


Our contributions are the following:

\begin{enumerate}
\item The first near full-disk model at 10-minute, 10-kilometer resolution forecasting geostationary cloud observations from 0--6~h that outperforms available baselines.

\item A demonstration that adding synoptic-scale input improves these forecasts, extending DOP to DOP+.
\end{enumerate}

We benchmark against a suite of available baselines including persistence forecasts, traditional synoptic-scale NWP, and pure DOP. For this domain, only synoptic scale global NWP is available. We trained a DOP model that uses the same base neural network as DOP+. We use DOP as a strong empirical baseline that has been shown to consistently improve upon more simplistic empirical methods such as optical flow \citep{dai_four-hour_2025, partio_cloudcasttotal_2025, pathak_learning_2026}.


\section{Results}

In this section, we describe the evaluation of DOP+. Our evaluation period ranged from June 13 to July 9, 2025 with daily initializations at 00, 06, 12, and 18 UTC. We also performed model size and timestep ablations which we show in the supplemental material.
The DOP and MPAS models are also evaluated for comparison.

\subsection{Quantitative evaluation}

Our measures of deterministic forecast skill are based on skill scores. These scores compare the forecast against a reference or random forecast. We first examine the mean absolute error skill score (MAESS) in order to assess the bulk skill of our forecasts. The reference forecast for our MAESS was the climatology: for the verification period, we calculated the per-gridpoint mean. Then we use the fractions skill score (FSS) to assess the skill of the models at different scales, and thus at the larger scales, alleviating the double penalty effect. Furthermore, we evaluate FSS using percentiles to reduce the effect of bias. As seen in Fig.~\ref{fig:maess_models}, MPAS has a systematic bias.

In addition, we evaluate based on a region bounded away from the western boundary. This avoids the impact of the fact that DOP does not have any information of boundary conditions. DOP on a global domain would not be disadvantaged by the fact that there are boundary conditions. The evaluation region was bounded in the west at $-100^\circ$ longitude, which is isolated from effects of the incoming flow over 12~h.

Evaluation focuses on Band 13, which resides in a relatively transparent atmospheric window with minimal water vapor absorption, providing the best estimate of cloud-top temperature. Cloud-top temperature is a robust indicator of cloud type and vertical extent, with colder temperatures associated with deep convective or high-altitude cirrus clouds and warmer temperatures indicative of low-level or no cloud. Other works have solely focused on Band 13 prediction. Compared to other ABI bands, it is the cleanest and sharpest observations we have of cloud evolution.

\subsection{Inter-model comparison}

\begin{figure}[t]
  \noindent\includegraphics[width=0.49\textwidth]{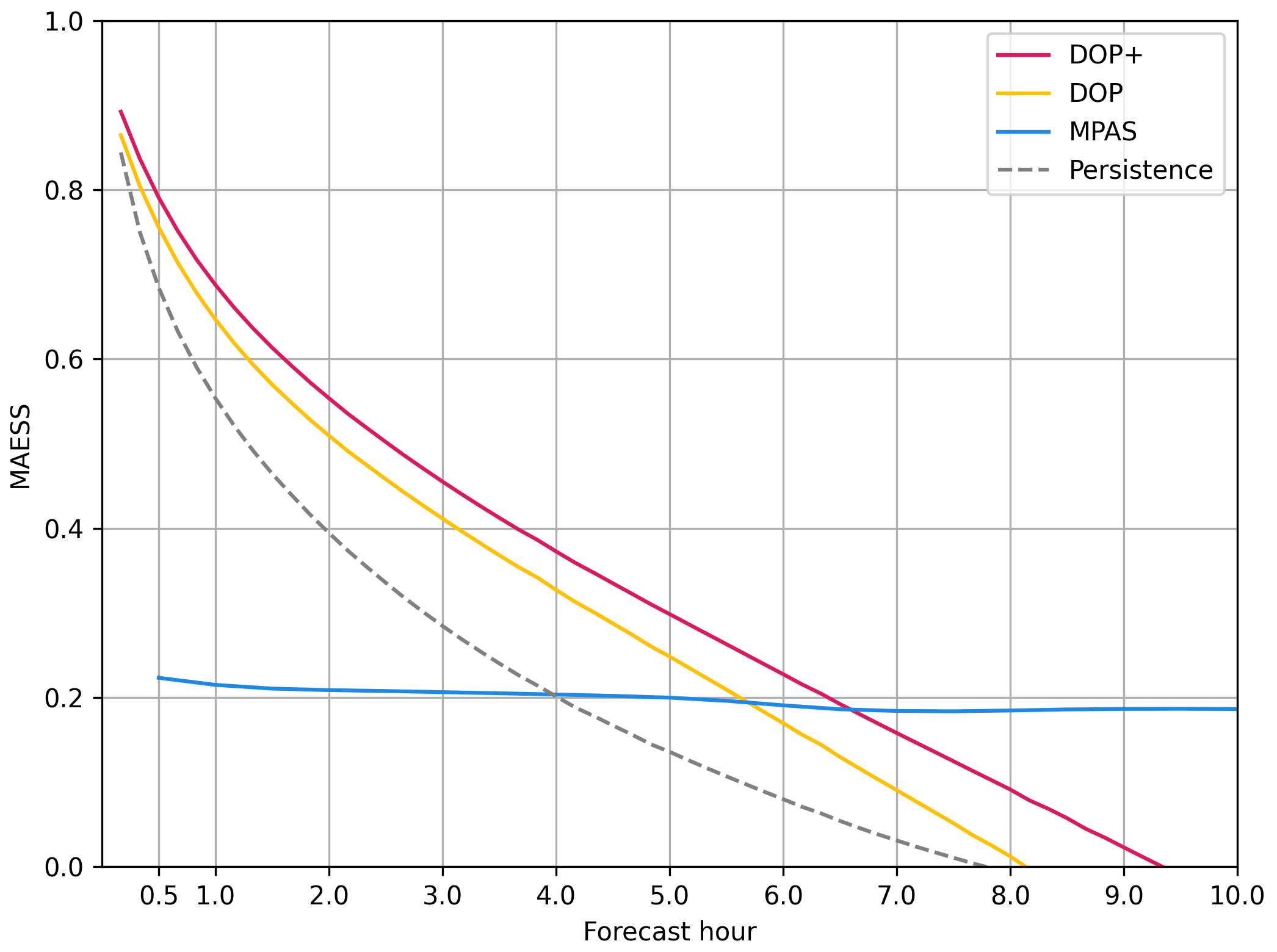}
  \includegraphics[width=0.49\textwidth]{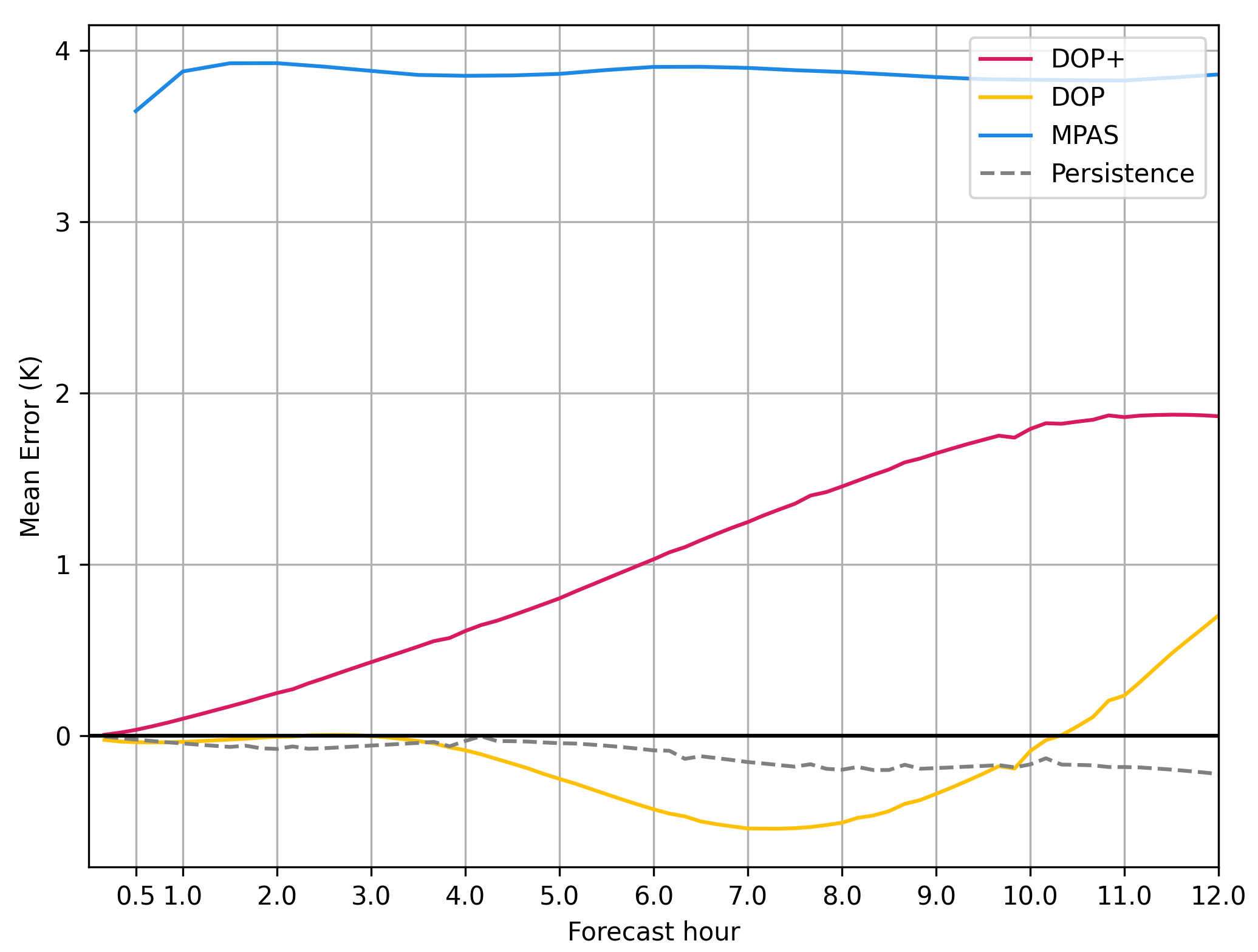}\\
  \caption{(left) MAESS and (right) Mean bias as a function of lead time for ABI band 13.}\label{fig:maess_models}
\end{figure}

Figure~\ref{fig:maess_models} shows that DOP+ outperforms MPAS in forecasting band 13, as measured by MAESS, up to around forecast hour 6, and outperforms DOP and persistence at all leads. For most lead times, DOP+ beats persistence by a margin of more than 0.1.  We have verified that the smallest improvements, of DOP+ over DOP, are statistically significant at the 0.95 level at all lead times.  Evaluation based on the percentile FSS yields similar results (not shown).

The additional skill of DOP+ relative to DOP and persistence indicates that DOP+ is gaining consequential information from the ERA5 fields. The ERA5 fields ingested by DOP+ are the resolved horizontal winds and thermodynamic state (see Table~\ref{tab:era5_vars}). These fields are relevant to the evolution of ABI brightness temperatures both directly, by informing the advection and propagation of cloud features, and indirectly, through the dependencies between coarse-grained cloud fields and other variables that form the basis for diagnostic cloud schemes in earlier generations of NWP models \citep{slingo_development_1987} and that learned parameterizations recover from data without condensate inputs \citep{gentine_could_2021}. Extending the evaluation of Fig.~\ref{fig:maess_models} to the entire domain gives further evidence of the importance of the ERA5 fields. The difference in performance between DOP+ and DOP is even larger when including the western boundary region, where there is generally flow into the domain and ERA5 estimates of evolving conditions at the boundary are particularly influential.

Despite this, the fact that the skill of DOP+ is notably worse than MPAS at longer lead times indicates that at these lead times, either DOP+ is not fully utilizing the ERA5 guidance or there is not enough information in the ERA5 fields. At longer leads, where the information from the satellite initial conditions has diminished, DOP+ should be relying on the ERA5 fields to predict the same satellite channels. We would therefore expect the skill of DOP+ to be comparable to that of MPAS.

Figure~\ref{fig:maess_models} shows a constant systematic bias in the MPAS predictions while the ML models have biases that grow over time. DOP also exhibits a cold bias during the forecast, resulting in a lower overall bias at 12 hour leadtime. It is encouraging that the bias of DOP+ seems to asymptote, indicating that it may have a better prediction of climatology than either DOP or MPAS.

\subsection{Fractions skill scores}

\begin{figure}[t]
  \noindent\includegraphics[width=\textwidth]{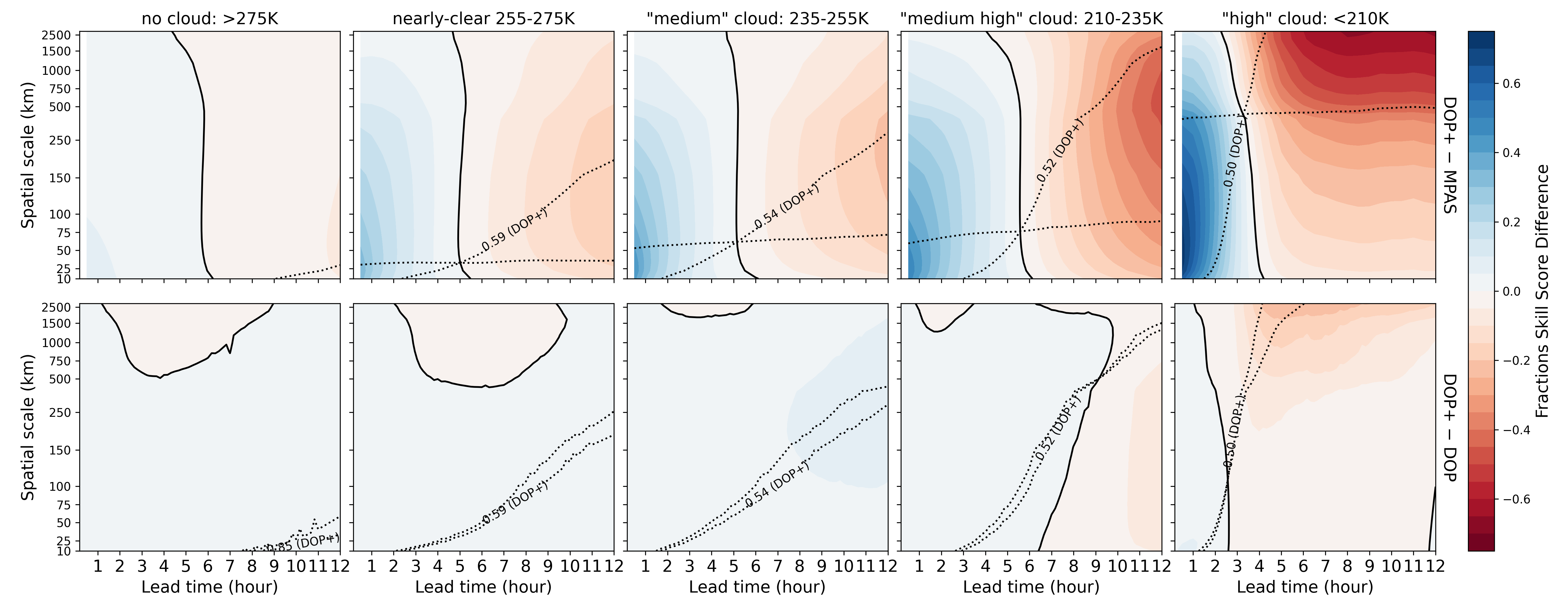}\\
  \caption{FSS difference of Band 13 over all five categories. (top) DOP+ minus MPAS and (bottom) DOP+ minus DOP. The dotted lines represent the contour of where each model in the comparison is skillful compared to a random baseline \citep{roberts_scale-selective_2008}. DOP+'s skillful contour is labeled while the unlabeled contour is the opposing model.}\label{fig:fss_diff}
\end{figure}

Our FSS categories correspond to thresholds from studies that roughly bins cloud types by their brightness temperature. ``High'' cloud is associated with overshooting tops \citep{bedka_objective_2010} and the other cloud categories align with the International Satellite Cloud Climatology Project definitions \citep{rossow_advances_1999}. ``Medium high'' is associated with deep convection, ``medium'' is associated with middle level clouds, ``nearly clear'' associated with low clouds, and clear associated with clear-sky conditions, though boundary layer clouds are also included in this category. The percentiles for each bin edge were computed by first determining the raw threshold's percentile in the observations, and then applying this percentile to the forecast.

We used percentile FSS to alleviate the effect of forecast bias. Instead of showing the raw FSS figures we show the difference in FSS between DOP+ and MPAS. We can thus more directly compare the models at various spatial scales, lead-times, and cloud categories. From Fig.~\ref{fig:fss_diff}, it is immediately apparent that from the smallest scales to the mesoscale, DOP+ is more skillful than MPAS, up to 5--6 hour lead time, except in the ``high'' cloud category. In general when looking left-to-right, it is apparent that DOP+ shows the most improvement over MPAS at earlier lead times, smaller scales, and higher cloud categories. This is consistent with the idea that DOP+ is predicting and ``assimilating'' what MPAS is bad at. MPAS-JEDI, a synoptic scale forecast system, has trouble assimilating small-scale, fast convective processes. DOP+ fills this gap without having to tune and operate a separate mesoscale prediction system such as the HRRR.

However, DOP+'s advantage decreases with increasing scale, leadtime and cloud-top height. In fact, MPAS has the largest advantage in performance over DOP+ in the ``high'' cloud category at lead times greater than 4 hours. The coldest categories correspond to deep convective cores and overshooting tops, whose realization depends on triggering processes that are subgrid at ERA5 resolution; these are the least predictable features and the ones for which the ERA5 guidance is least informative. As noted above, MPAS skill should be a rough lower bound on DOP+, and here it is not.


DOP+ and DOP are very close in FSS relative to MPAS. Again, DOP+ is superior to DOP by a small margin. At longer lead times, in the ``high'' cloud category, the skill of DOP is superior to DOP+. This is also exhibited to a lesser extent in the ``medium-high'' category. We are unsure of the reasons for this but suspect that it could be because of DOP's tendency to grow and intensify large cold cells more intensely than DOP+. Further analysis is needed to determine the true reasons for this. Additionally, previous work has also shown that adding information to the ML model's input does not clearly improve forecasts for all metrics simultaneously---some metrics may actually degrade with more information \citep{miralles_pointwise_2026}. It is possible that the large-scale guidance is confusing the model at these longer lead times, for these metrics.

\begin{figure}[t]
  \noindent\includegraphics[width=0.49\textwidth]{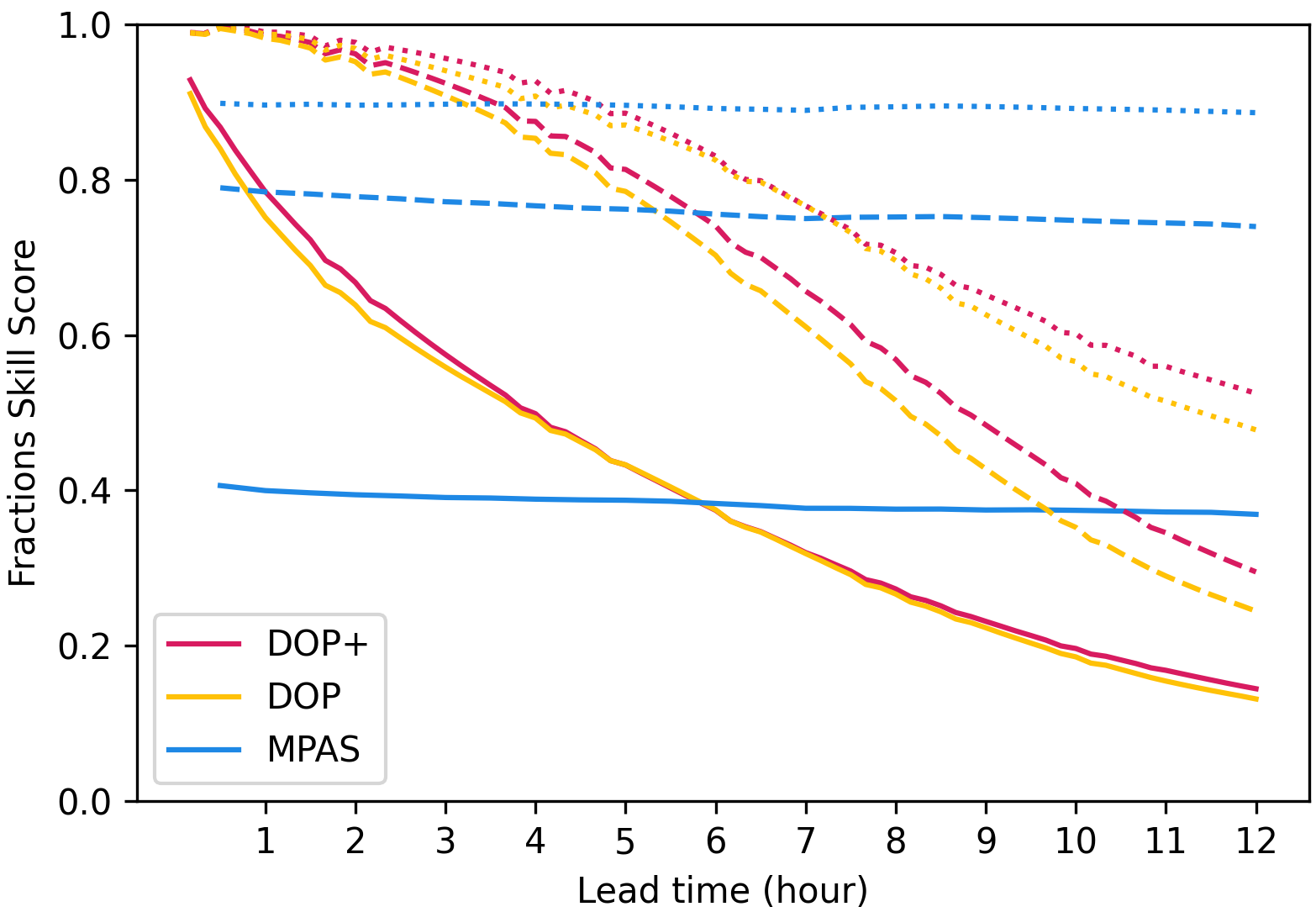}
  \includegraphics[width=0.49\textwidth]{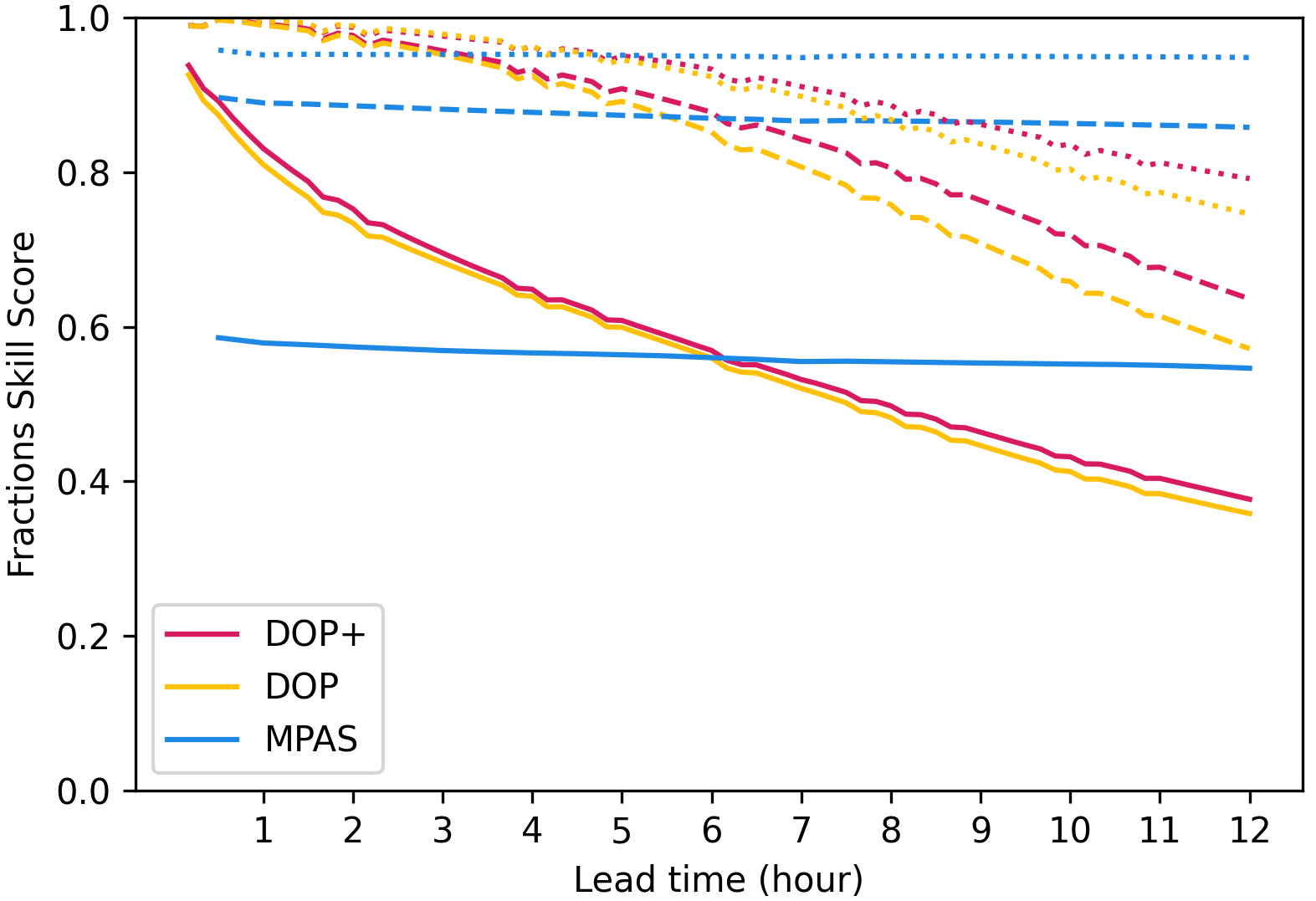}\\
  \caption{FSS of band 13 comparing the 3 models (left) 235~K threshold (right) 255~K threshold. Different line styles represent different FSS pooling windows: solid: 11~km grid scale; dashed: 440~km mesoscale; dotted: 2200~km synoptic-scale.}\label{fig:fss_thresholds}
\end{figure}

Figure~\ref{fig:fss_thresholds} shows percentile FSS at two cloud-relevant thresholds of 235~K (deep convection) and 255~K (medium cloud) rather than bins. DOP+ shows superior performance against MPAS at lead times out to 6 hours at the smallest scale. The advantage of DOP+ however decreases at the larger FSS spatial scales. The crossover point is roughly 4.5 hours at the synoptic scale.
DOP shows similar performance at earlier lead times across scales, with DOP's skill degrading relative to DOP+ at later lead times. The largest difference between DOP and DOP+ is at the mesoscale.

As argued in previously, MPAS skill should act as a lower bound on DOP+ at long leads: there should be no dark red in the FSS difference figures, and no crossover in the threshold figures. Instead, at longer leads and larger scales, DOP+ falls below MPAS.

\begin{figure}[t]
  \noindent\includegraphics[width=\textwidth]{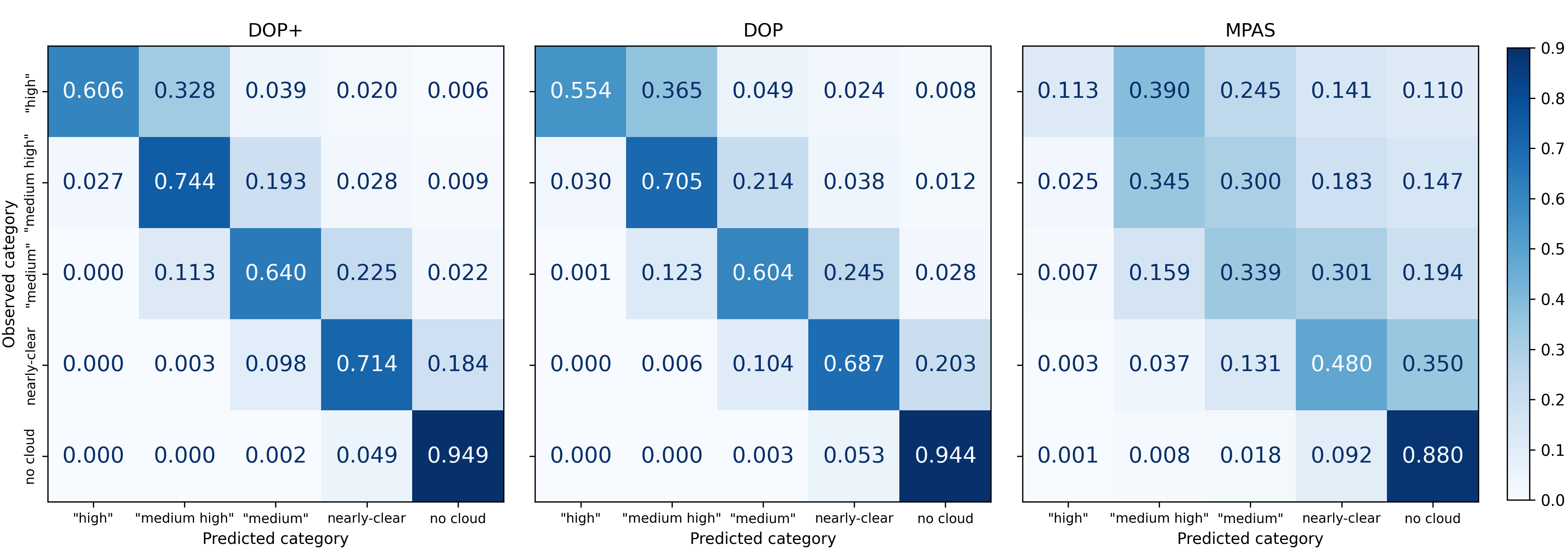}\\
  \caption{Row-normalized confusion matrix at 1~h lead time for the different cloud categories for (left) DOP+, (center) DOP, and (right) MPAS. As with the FSS, these were computed with percentile bins.}\label{fig:confusion_models}
\end{figure}

Figure~\ref{fig:confusion_models} shows row-normalized contingency matrices, computed using percentile thresholds to alleviate the effect of bias. The figures shown are for 1 h leadtime forecasts. MPAS is warm-biased, with significant prediction errors 2 categories away. DOP+ and DOP matrices are dominated by the diagonal, with mispredictions mainly only 1 category away. Overall, both ML methods tend to mispredict by predicting too warm rather than too cold. DOP+'s advantage over both baselines increases as the brightness temperature decreases, where the clouds generally become harder to predict. MPAS struggles the most at colder clouds, while the ML methods have more uniform prediction performance across categories.

\subsection{Forecast smoothing}

\begin{figure}[t]
  \noindent\includegraphics[width=\textwidth]{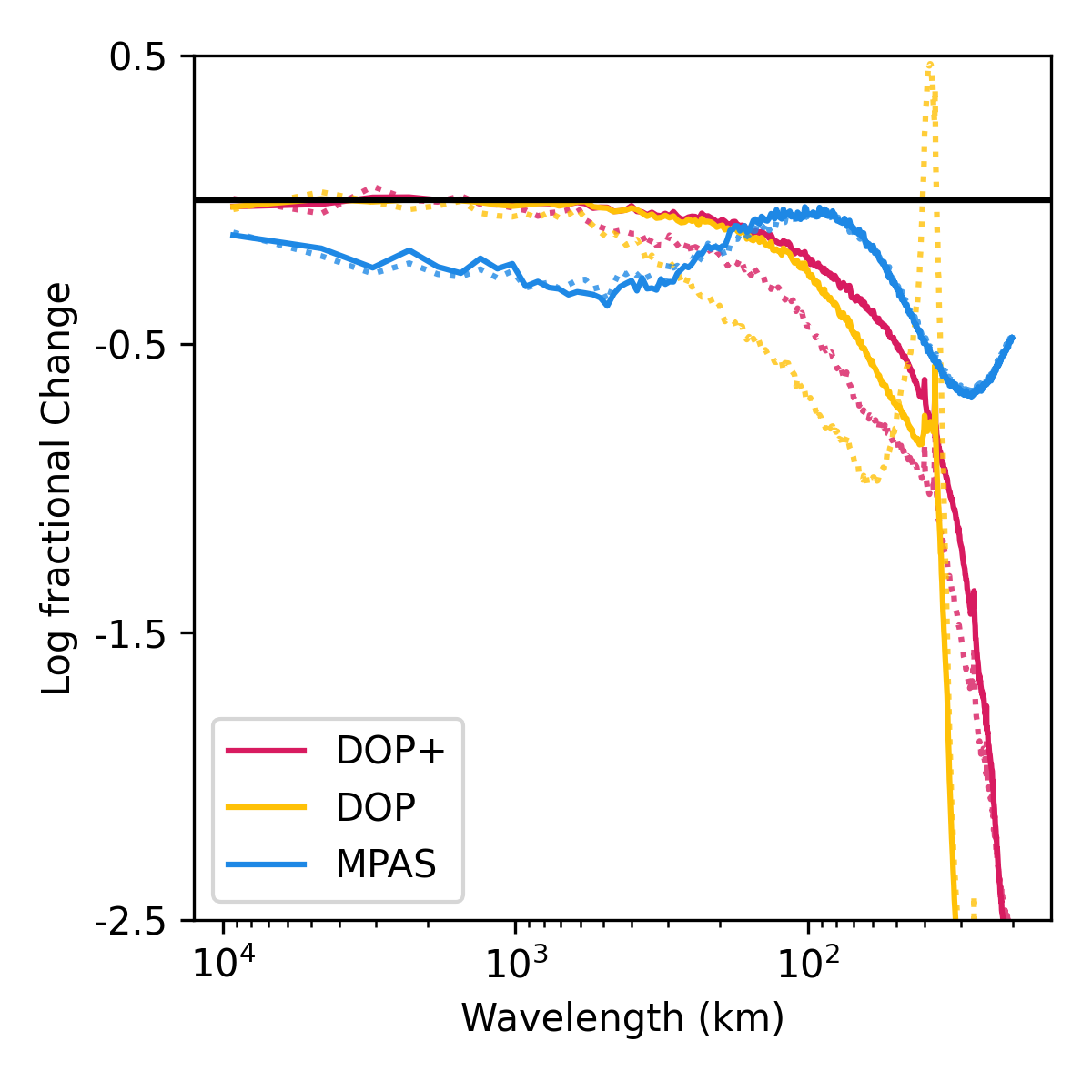}\\
  \caption{Log ratio of power spectral density (PSD) between forecast and reference (truth) fields for Band 13, as a function of spatial scale, for DOP+, DOP, and MPAS at 1-hour (solid) and 3-hour (dotted) lead times. A value of zero indicates that the forecast reproduces the spectral power of the reference at that scale. Negative values indicate smoothing at that scale and positive peaks indicate spurious noise.}\label{fig:psd}
\end{figure}

Figure~\ref{fig:psd} shows the log ratio of power spectral density (PSD) between the model predictions and the observations at 1 and 3 h lead times. At large spatial scales ($>$1000~km), all models closely track the reference, with log PSD ratios near zero, indicating that synoptic-scale structures are well represented at both lead times. MPAS shows no significant variation in spectra with lead time. As expected with deterministic, MSE trained models, the ML methods smooth more at larger lead times due to autoregressive smoothing compounding over successive rollout steps. The smoothing becomes significant at spatial scales less than 100~km. However, for the first forecast hour, DOP+ and MPAS have comparable spectra at scales above 20~km ($2\Delta x$ or near grid scale).

This indicates underrepresentation of small-scale features in DOP+ at 3-h relative to the observations, and thus at longer lead times has coarser effective resolution than MPAS. The spectra at longer leads likewise point to issues in using the ERA5 guidance.

Overall, DOP tends to smooth more than DOP+ at all lead times. But at around the 50~km scale, DOP also injects high-frequency noise. The peak in the power spectra is apparent at 1~h lead time and then grows to a large peak at 3~h lead time. These effects are replicated across training ablation experiments for the same timestep and architecture. Thus DOP both smooths more and injects more spurious noise. Using multistep training is the way to mitigate this noise but, doing so would increase smoothing and decrease DOP's ability to represent fast phenomena. Thus, since the focus here is to represent fast phenomena, we chose to not multistep train any of the models.

\subsection{Inter-channel comparison}

\begin{figure}[t]
  \noindent\includegraphics[width=\textwidth]{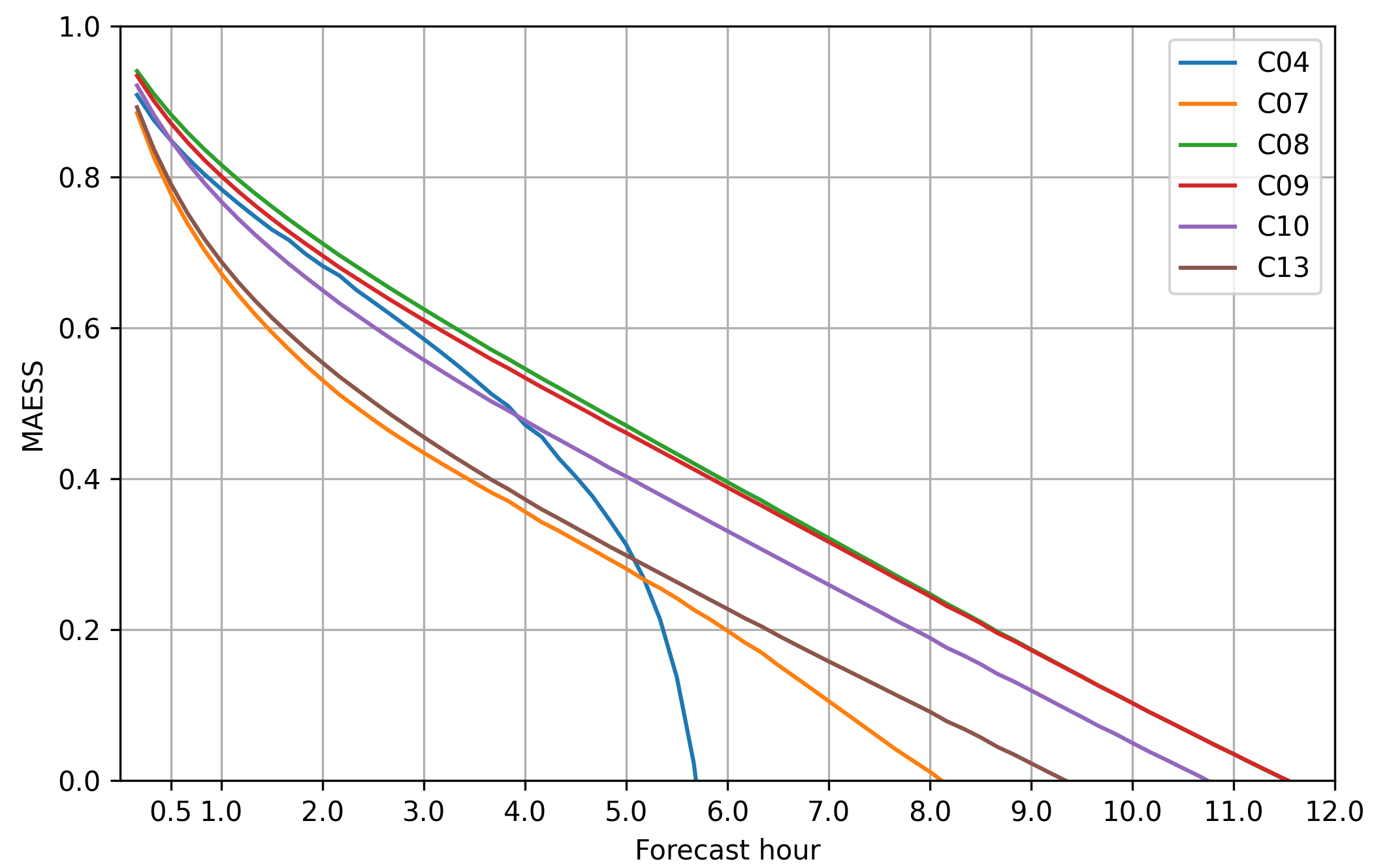}\\
  \caption{MAESS of DOP+ as a function of lead time for each individual ABI band.}\label{fig:maess_channel}
\end{figure}

Figure~\ref{fig:maess_channel} shows the MAESS of DOP+ forecasts for each channel. For the IR channels, DOP+ is skillful for 8 hours, and even longer for other channels. Among the IR channels, band 7 and 13 show the least skill. This is expected because bands 8--10 are water vapor bands which evolve at much larger time and space scales than bands 7 and 13. The differences between bands 8--10 are also similarly explained by the variation in time and space scales: band 10 is low-troposphere water vapor, 9 is mid-troposphere, and 8 is upper-troposphere. Unlike band 13, band 7 is sensitive to solar reflectance and is more sensitive than band 13 to the hottest part of the pixel. Thus it is unsurprising that band 7 has degraded skill compared to band 13, especially at longer lead times.

The outlier on this figure is band 4 ``cirrus band'' which is attributable, like band 7 to the sensitivity to solar reflectance. When evaluating on the full prediction region, band 4's performance was worse. This is due to the dependence of band 4 on solar reflectance and its primary function of detecting high cirrus clouds. On the western boundary, DOP+ needs to fully rely on the ERA5 fields to advect-in cirrus clouds. Thus either DOP+ is not utilizing the ERA5 information for the cirrus clouds or since cirrus clouds are high and difficult to represent in NWP, the ERA5 fields do not contain enough information for DOP+ to represent the cirrus clouds. Finally, DOP+ could be paying insufficient attention to the TSI field (top of atmosphere total solar irradiance), thus mispredicting the visible band due to the diurnal cycle.

\subsection{Qualitative evaluation}

We evaluate by looking at two cases.

\begin{figure}[H]
  \noindent\includegraphics[width=\textwidth,height=0.85\textheight,keepaspectratio]{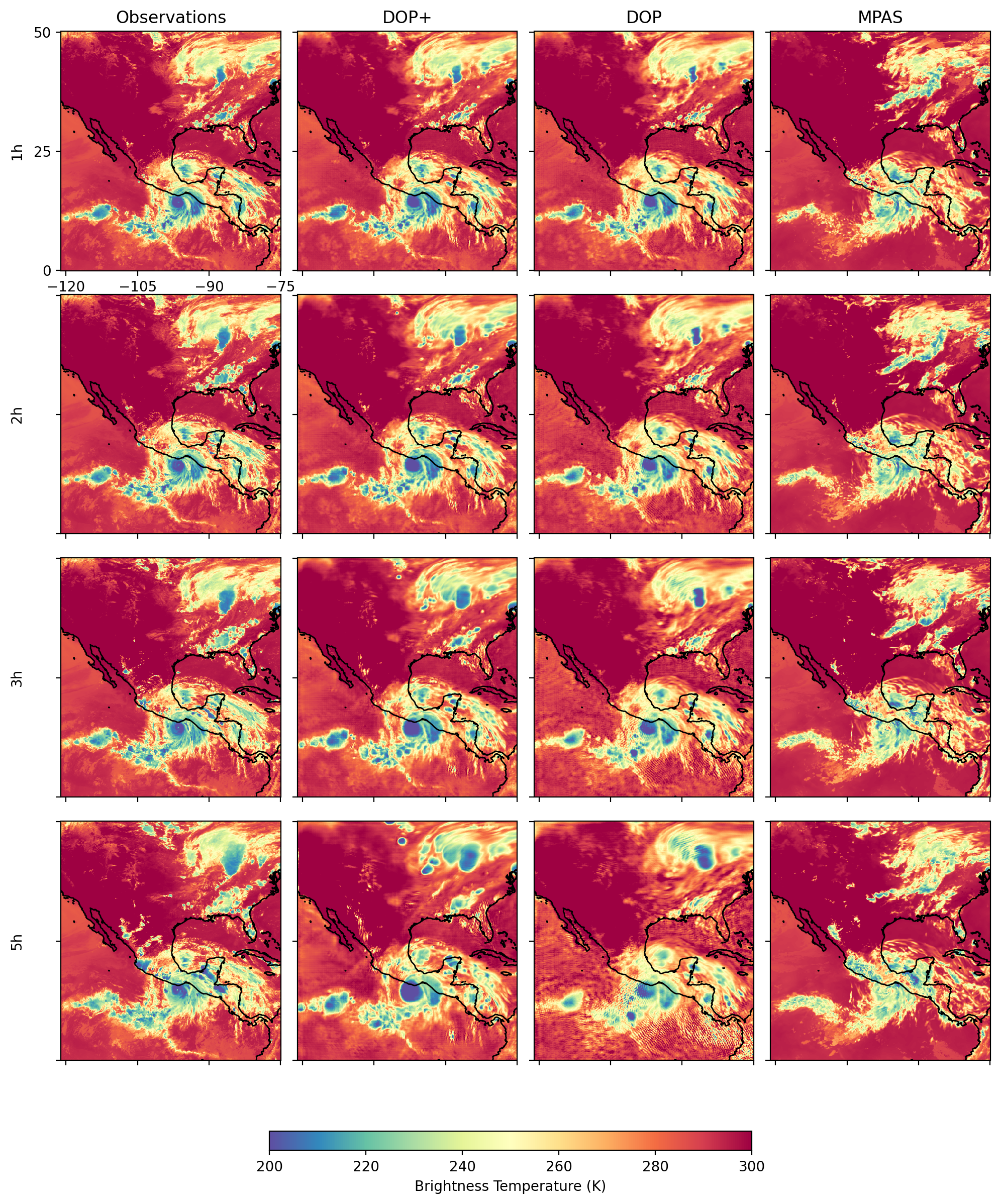}\\
  \caption{1, 2, 3, and 5 hour forecasts initialized at 1800 UTC 18 June 2025 over a subdomain of the model's domain. Several large organized convective systems are represented. Hurricane Erick, a category 4 hurricane, is off the coast of Mexico. A mesoscale convective system (MCS) is over the eastern US. The MCS intensifies as the forecast progresses.}\label{fig:case1}
\end{figure}

Figure~\ref{fig:case1} shows forecasts initialized at 06:00 UTC on 18 June 2025 at lead times of 1, 2, 3, and 5~h, alongside the verifying observations. The case contains two organized systems of particular interest: a hurricane in the eastern Pacific, and a mesoscale convective system (MCS) over the eastern United States. Both ML models represent the hurricane's cold central cloud shield and position throughout the forecast, though its structure smooths with lead time. MPAS does not have a cold core. The MCS is propagated coherently by the ML models out to 5~h with more accurate intensity than MPAS. Though no model captures its full observed intensity, 5~h is a lead time at which such systems are typically beyond the reach of pure extrapolation methods.

DOP, compared to DOP+, tends to both smooth the field by combining distinct cloud features and injecting unrealistic high-frequency noise features. These high frequency wave-like cloud structures become more apparent at later lead times. In this particular case DOP+ overpredicts cold features, separate from the MCS, in central CONUS compared to the other two models.

DOP+ grows what could be small seeds of convection to the west of the MCS. However it is unclear whether this convective growth is realistic for the physical conditions since they do not appear in either MPAS or the observations.

The ML models do not capture the convective initiation over northern Florida. They also do not capture the cloud formation in northern Mexico. MPAS exhibits these features albeit with lack of accuracy in placement and intensity, corresponding to the predictability of these features. These limitations of DOP+ point again to issues with the information usage from the ERA5 input.

\begin{figure}[H]
  \noindent\includegraphics[width=\textwidth,height=0.85\textheight,keepaspectratio]{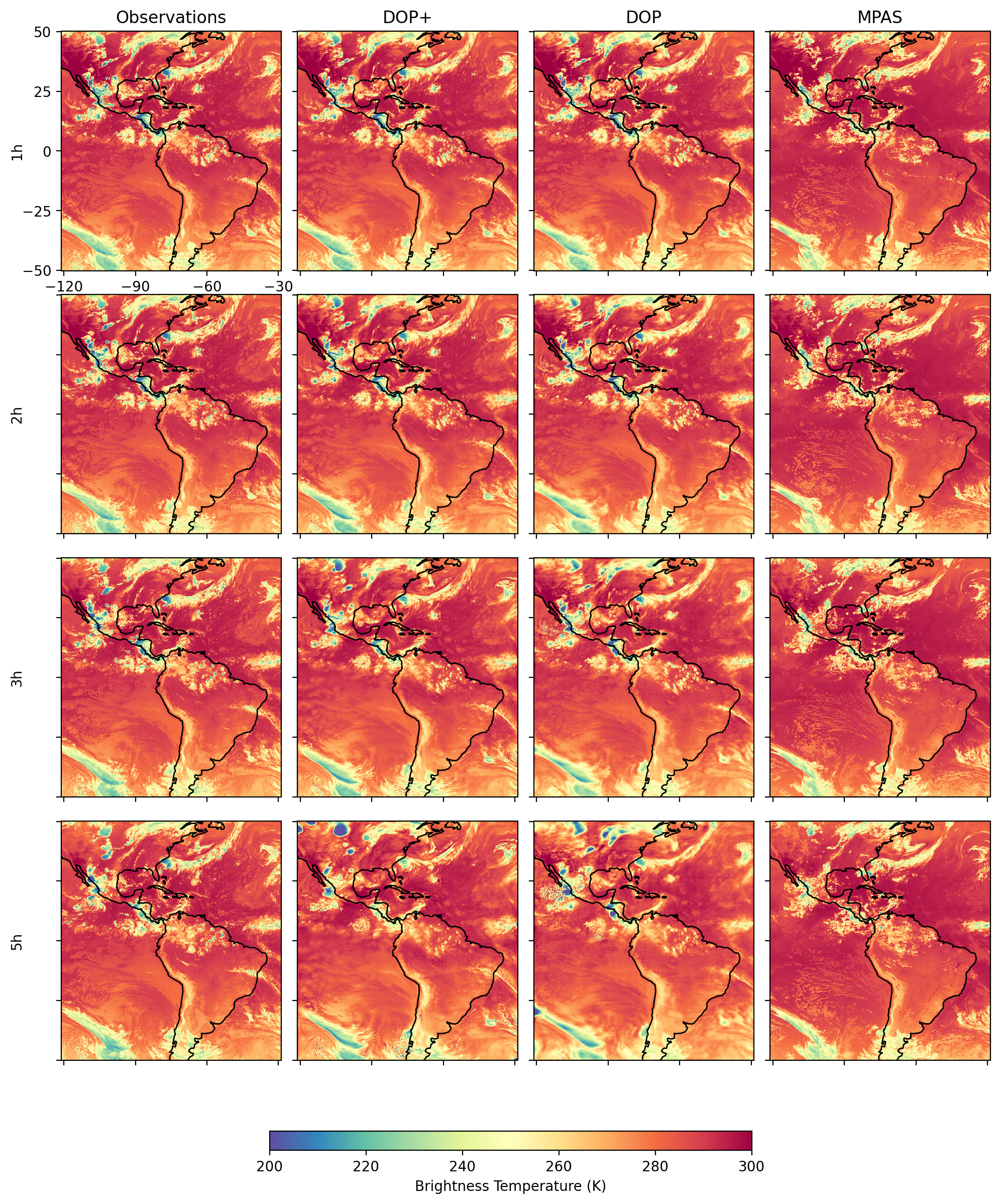}\\
  \caption{1, 2, 3, and 5 hour forecasts initialized at 0000 UTC 6 July 2025. A typical N.\ hemisphere summertime case is illustrated. Isolated cloud formations develop, advect, and dissipate over CONUS and the tropics. A large frontal system is advected in the southern ocean with parts advected in from the boundary.}\label{fig:case2}
\end{figure}

We now discuss a case in Fig.~\ref{fig:case2} without large organized convective systems. At short lead times both ML models closely reproduce the overall cloud field, but their behaviors diverge as the initial condition signal decays. The distinct difference is in the western boundary: DOP has no knowledge of conditions outside the domain, and by 3 and 5~h large spurious cold features intrude along the western boundary near 10$^\circ$N and 25$^\circ$S, where the flow enters the domain. DOP+ largely suppresses these artifacts, consistent with the synoptic-scale input supplying the inflow information that the observation history cannot. Both ML models, however, share a comparable overprediction of cold brightness temperatures, with cold cloud features growing and persisting beyond their observed extent indicating that this deficiency is not remedied by synoptic conditioning alone. At 5~h leadtime, intense cold clouds appear in the southern Atlantic where it is winter. This indicates that the ML prediction is not 100\% physically constrained by the ERA5 input.

The two models also differ in texture: DOP produces some visibly smoother fields than DOP+ at longer leads, while simultaneously exhibiting high-frequency noise over equatorial South America, again visible here as speckled fine-scale structure absent from the observations.

MPAS misses much of the marine boundary-layer cloud in the southeastern Pacific off the South American coast.
In contrast, both ML models have significantly better representations of these clouds.
MPAS also again substantially underpredicts the intensity of cold cloud tops throughout the domain. 
Like in both the hurricane and the MCS, convective features appear warmer and more diffuse than observed.
These qualitative failure modes: boundary-layer cloud, convective intensity, and organized systems, are the regimes in which the parametrizations of synoptic-scale NWP are known to struggle, and in which the direct-observation approach shows a large advantage.

\section{Discussion}\label{sec:discussion}


We have presented a near full-disk model that forecasts GOES-East ABI observations at 10-minute, 10 km resolution, outperforming DOP, persistence, and MPAS. DOP+ outperforms MPAS at up to 6~h lead time at convective- and meso-scales. The model predicts evolution of fast processes directly in observation space with an unprecedented combination of domain size and temporal resolution. Convective structure is preserved out to 2--3~h. We show that conditioning on large-scale meteorological guidance improves these forecasts---extending direct observation prediction to DOP+. The domain, spanning tropics and midlatitudes, land and ocean, and both hemispheres' seasons at once, requires much of the regime diversity a global system would face. Thus we lay the foundation for truly global high-resolution prediction with DOP+.

Our work is a meaningful step toward global prediction in the demands it places on a single model. Within our GOES-East domain, one model must represent deep tropical convection and midlatitude baroclinic development, continental land surfaces (CONUS, Amazon) and vast ocean basins, across both hemispheres' seasons simultaneously. It must predict some of the atmosphere's least predictable phenomena such as mesoscale convective systems, while also representing the slow synoptic flow of the Southern Ocean. A limited-area model does not reconcile these regimes within a single set of parameters.
Our model attains skill across this range, spanning both fast and slow processes, without regime-specific tuning.
This we argue, is a substantive result: the difficulty of the global problem is largely the difficulty of regime diversity, and much of that diversity is already present here. This gives us confidence that the approach will extend to the high latitudes, the regime still missing from our demonstration.


In contrast to MPAS, DOP+ and DOP are quite similar both quantitatively and qualitatively. Hence the value-add of the ERA5 guidance in DOP+ is minor compared to the signal from the observations. Consequently, we expect switching the model guidance from ERA5 to a true NWP forecast to have minimal effect. This is further supported by reasoning outlined in Sec.~\ref{sec:methods}: any information gained from the guidance fields will be purely large-scale, and synoptic scale forecast skill does not degrade much at these scales over 0--12~h lead times. Future work could explore fine-tuning using real forecasts and evaluating with real forecasts as guidance. We view this direction as lower priority than the questions we turn to next.

At longer lead times, a conclusion throughout our results is that DOP+ has issues with using the synoptic-scale ERA5 input. Several pieces of evidence point this way:

\begin{itemize}
\item Qualitative evaluation indicates that DOP+ misses some convective initiation, especially at longer lead times. For certain convective cells, MPAS shows convective initiation while DOP+ does not. This is further supported by the fact that the FSS, at cold brightness temperatures, of DOP+ is worse than MPAS at these longer lead times. At these longer leads, DOP+ should be relying on large-scale guidance for convective conditions.

\item There is qualitative cold bias associated in round cloud features, including convection that DOP+ does initiate. These tend to grow and fail to dissipate over the forecast, driving brightness temperatures colder than observed. These overly large, homogenous cold and round features are not represented in the training data, thus the model does not learn how to dissipate them.

\item The model occasionally produces isolated, very cold, convection-like features in the winter hemisphere, in regimes where the synoptic input clearly indicates stability. These are a signature of a model applying a learned convective-evolution prior without conditioning on physical laws.

\item Channel 4 performance degrades significantly near the domain's western boundary where the flow comes in.

\item Most tellingly, our forecast skill degrades below that of MPAS at long leads. A model that sufficiently relies on the NWP input should, in the limit, be able to fall back on it and match it. That the skill of DOP+ crosses below NWP suggests the guidance information is not available to the model in a form it can lean on when the observational signal has decayed.
\end{itemize}

Though the ERA5 fields are likely insufficient for some cloud types, we suspect the dominant mechanism is our training objective. We train on a single 10-minute step to best represent fast processes. For this objective, the best available predictor of the satellite field is overwhelmingly the previous satellite field. Advection and short-term persistence explain nearly all the variance the loss can see. The synoptic input offers, at that horizon, a marginal improvement over an already-excellent predictor, so gradients pushing the model to use it are small. The model is behaving rationally with respect to the objective we gave it. The objective does not incentivize the behavior we want at the longer lead times.

This suggests the fix is to train at longer leads, where the satellite initial  conditions' signal has decayed and the model must rely on the synoptic fields to do better than climatology. Multi-step training would incentivize this directly.
But deterministic multi-step training has a well-known effect: minimizing a pointwise loss over mean-squared error incentivizes the conditional mean, and the conditional mean of low-predictability processes, like convection, is overly smooth \citep{subich_fixing_2025}. We would lose the sharp, convective-scale structure that motivated our approach in the first place.

Fast processes, like convection, are the most difficult phenomena our model is asked to predict, in large part because convection is poorly observed by the ABI instrument. Band 13 sees only cloud top; other bands offer coarse moisture information, and prior work suggests the ABI channels have retrievable temperature profile information \citep{schmit_legacy_2019}. But none of this is a direct view of the boundary-layer moisture, convergence, or vertical structure of instability that determines whether convection initiates. The ERA5 guidance we provide is likewise too coarse to supply this information at the scales we operate at.

Hence we cannot cleanly attribute the model's failure to any one cause. The relevant information may simply be absent from the observations. It may be present in the ERA5 input but not transferred to the model, for the training-objective reasons discussed above. Or it may be absent from the ERA5 input itself, since it is a synoptic-scale product that parametrizes convection rather than resolving it. Most likely the truth is some combination, and our current experiments do not separate these possibilities. Ablations that vary large-scale guidance input quality independently of the training horizon would help diagnose which is dominant.

These threads converge on an approach already established in the MLWP literature: stochastic modeling. It is highly warranted: convective evolution, given our current observations, is not deterministically predictable. A deterministic ML model for these processes is limited: as explained above, hard-to-predict features will be overly-smoothed.
Though deterministic training is a limitation, it is not a disqualifying one as our results establish where a direct-observation approach stands against operational NWP.
The forecasts represent small-scale information out to 2--3 hours and holding the objective fixed across DOP and DOP+ reveals the effect of synoptic guidance.

A stochastic formulation could enable substantial improvements. Because a probabilistic objective such as CRPS is not minimized by the conditional mean, it does not penalize spatial sharpness, so multi-step training under it need not smooth. This lets us train at the longer leads where the model must attend to the synoptic fields, and can fully use them as probabilistic conditioning. This change could address predictability, smoothing, and NWP integration together. Additional observations, from more ABI channels or other instruments, would help further. We will take on all these steps in a follow-on stochastic DOP+ model.

Our current work also makes a stochastic model more tractable. Architecture, input configuration, and implementation are far cheaper to iterate on in the deterministic setting, where training is faster and forecast errors are easier to diagnose. These choices then carry over: prior work in MLWP has shown that a deterministic model can be converted into a skillful ensemble by fine-tuning on a probabilistic objective such as the continuous ranked probability score \citep{schreck_controllable_2025, zhong_fuxi-ens_2025}.

We leave the inclusion of the high latitudes to future work, and we expect it to require substantially more effort than simply extending the domain. Previous work has demonstrated skillful forecasting of diagnosed total cloud cover over a small region of northern Europe (50$^\circ$--75$^\circ$N) \citep{partio_cloudcasttotal_2025}, so the problem is not intractable in principle.

At these high latitudes, we face observational constraints. The coverage of high-quality geostationary imagery ($<$~74$^\circ$ local zenith angle) falls away poleward. The imagery at these high latitudes, with high LZAs, is distorted by the viewing geometry and could be unreliable for the fast processes we target. Closing this gap will require observations from other platforms such as polar orbiters, whose observation density is highest where the geostationary observations are the sparsest. Polar orbiter observations are sparse and irregular in time and space, and assimilating them into an autoregressive ML forecast system significantly increases data-handling and architectural complexity. For this demonstration we chose to abstract away from that complexity and work within a high-quality region of ABI observations to model fast processes.

We do not have a clean explanation why the percentile binned FSS curves for MPAS change only slightly with lead time. One hypothesis is that MPAS predictions are not tracking the fine-scale details even at short leads. Due to its biases and deficiencies in data assimilation, its convective-scale detail is sufficiently decorrelated from the truth, and its FSS therefore reflects its ability to reproduce the statistics of cloud fields at a given scale rather than their placement. If so, there is no phase information left to lose, and the curve is flat because it is already at its asymptote. \citet{pathak_learning_2026} report similar behavior for the HRRR, which is at least consistent with our results. We flag this as an open question.


\section{Methods}\label{sec:methods}

\subsection{Datasets}

GOES-East (GOES-16/19) is a GOES-R series geostationary satellite operated by NOAA and located above 0$^\circ$ latitude, 75.2$^\circ$W longitude. It provides full-disk imagery every 10 minutes in its standard Mode 6 scanning operation and has a native spatial resolution at nadir ranging from 0.5~km (visible) to 2~km (infrared) depending on the channel.

The Advanced Baseline Imager (ABI), an instrument onboard the GOES-R series, provides data in 16 spectral bands spanning the visible, near-infrared, and infrared portions of the spectrum \citep{schmit_closer_2017}. In this study, six ABI channels are used: Bands 4, 7, 8, 9, 10, and 13 (see Table~\ref{tab:abi_bands} for details). These channels were selected for their complementary sensitivity to cloud-top properties, cloud vertical structure, and atmospheric moisture at multiple tropospheric levels. Band 4 detects thin cirrus, Band 7 provides sensitivity to low-level stratus, fog, and cloud particle phase, Bands 8--10 capture water vapor signals at upper, mid, and lower tropospheric levels respectively, and Band 13 provides a stable longwave brightness temperature (BT) less affected by water vapor absorption, serving as the primary proxy for cloud-top temperature or in clear-sky conditions, surface skin temperature.

\begin{table}[t]
\caption{DOP+ and DOP input and output ABI bands.}\label{tab:abi_bands}
\centering
\renewcommand{\arraystretch}{1.25}
\begin{tabular}{ll p{4cm} p{6cm}}
\hline
ABI Band & Wavelength ($\mu$m) & Name & Description \\
\hline
4           & 1.378              & Cirrus              & Near-IR channel for daytime detection of thin cirrus clouds \\
7           & 3.9                & Shortwave IR Window & Detects low clouds, fog, and fire hotspots; usable day and night \\
8, 9, 10    & 6.19, 6.95, 7.34  & Water Vapor (Upper, Mid, Lower) & Upper-, mid-, and lower-tropospheric water vapor tracking, jet stream analysis, and moisture estimation \\
13          & 10.35              & ``Clean'' Longwave IR Window & Less sensitive to water vapor than other IR window channels; used in many composite products and cloud analyses \\
\hline
\end{tabular}
\end{table}

The native-resolution channel data were coarse-grained to a regular 0.1$^\circ$ grid (approximately 11~km at the equator) by spatial averaging, reducing computational costs while retaining the cloud features of interest. The spatial domain was first restricted to retain only pixels with a local zenith angle (LZA) less than 74$^\circ$, which is the region used by the operational ECMWF IFS system \citep{burrows_assimilation_2020}. Then, the largest rectangular geographic domain (by approximating a square) fully contained within the LZA $<$ 74$^\circ$ region was then subsetted, yielding a region spanning latitudes 50$^\circ$S to 50$^\circ$N and longitudes 121$^\circ$W to 29$^\circ$W. This domain encompasses the tropical and subtropical Americas, the Caribbean, and the adjacent Atlantic and eastern Pacific Ocean basins.

ABI Level 1b radiance data were converted to brightness temperatures (for IR channels, Bands 7--13) and reflectance factor (for the near-IR channel, Band 4) following the GOES-R Product Definition and Users' Guide \citep{noaanasa_goes-r_2019}. Any time step in which more than 25\% of pixels within the domain contained missing or fill values was discarded entirely. Channel 4 was scaled with a log-normal transform, and all other channels with a standard-normal transform. For the remaining timesteps, residual missing or fill values were replaced with channel-specific climatological means (by filling with 0 after the transform).

Training data from April 2018 to July 2022, and validation data from July to December 2022 (totaling 165632 samples) were drawn from the NOAA reprocessed ABI Level 1b archive, which incorporates artifact corrections across all channels, plus radiometric calibration improvements for channels 1--7  \citep{noaanasa_goes-r_2024}. Evaluation data from June--July 2025 were sourced from the operational Level 1b archive hosted on AWS, reflecting the current operational calibration without reprocessing.

\subsection{ERA5 and static fields}

\begin{table}[t]
\caption{DOP+ and DOP input variables and ERA5 levels.}\label{tab:era5_vars}
\centering
\begin{tabular}{l c}
\hline
 Variables  & Levels$^{a}$  \\
\hline
 \cline{2-2}
Atmospheric Variables:  & \multicolumn{1}{|c|}{500 hPa,}\\
  ~~~$U$ component of wind  & \multicolumn{1}{|c|}{70 ($\sim$165 hPa),}    \\
  ~~~$V$ component of wind  & \multicolumn{1}{|c|}{80, 90, 95, 100,}\\
  ~~~Temperature &   \multicolumn{1}{|c|}{110, 120, 130, 105,} \\
  ~~~Specific humidity  &   \multicolumn{1}{|c|}{136, 137 (bottom)}   \\
\cline{2-2}
\hline
Single-level Variables: & \\
    ~~~Geopotential Height (Z) & 500 hPa \\
    ~~~Total solar irradiance & Top of atmosphere \\
    ~~~2-m Temperature & Surface \\
    ~~~Surface Pressure&  -\\
\hline
Static Variables: & \\
  ~~~Surface Geopotential & Surface\\
  ~~~Land Mask & - \\
\hline
\multicolumn{2}{l}{$^{a}$Unitless integers represent model level indices.}
\end{tabular}
\end{table}

For model guidance, the ERA5 reanalysis dataset was used \citep{hersbach_era5_2020}. ERA5 has a 0.25$^\circ$ spatial resolution, with hourly global coverage. For this study, we subsetted 11 model levels from the lowest model level up to the stratosphere. Other static and dynamic variables were also included (see Table~\ref{tab:era5_vars}). The data was then fine-grained to the same 0.1$^\circ$ grid as the ABI data with bilinear interpolation. All variables were standard-scaled. High-resolution terrain characteristics were coarse grained from the USGS 30 arc-second (${\sim}$1~km) global multi-resolution terrain elevation data dataset \citep{danielson_global_2011}.

Though ERA5 is not a forecast product, using it as the synoptic guidance drastically simplifies training and data management. Two considerations make this a reasonable choice. First, synoptic-scale forecast skill degrades only modestly at 0--12~h lead times, so a real forecast field would closely resemble the corresponding analysis over our forecast window. Second, synoptic products have an effective resolution of more than 200~km \citep{bolgiani_wind_2022}, an order of magnitude coarser than our satellite input. Any information the model gains from the guidance is therefore purely large-scale. We note that ERA5 nonetheless has smaller errors relative to the observations than a forecast would, so our results should be read as a slightly optimistic estimate of the benefit of synoptic conditioning. Verification with real forecast output, from either NWP or MLWP, is left to future work. We discuss this choice in the discussion.

\subsection{MPAS baseline}

For the synoptic-scale NWP baseline we used the Model for Prediction Across Scales--Atmosphere \citep[MPAS-A;][]{skamarock_multiscale_2012} on a quasi-uniform 15-km mesh with a standard ``mesoscale reference'' physics suite, coupled to the Joint Effort for Data assimilation Integration \citep[JEDI;][]{tremolet_jedi_2020}. A global, synoptic-scale model is the appropriate baseline because we do not yet have higher-resolution operational forecast systems on a domain as large as ours, and MPAS-JEDI provides a robust physics-based global NWP framework \citep{liu_data_2022, guerrette_data_2023, ban_mpas_2026}.

The verification period is based on a month-long, six-hourly MPAS-JEDI cycling experiment spanning 10 June to 9 July 2025, of which the first three days were discarded as spin-up. Beginning on 13 June 2025, forecasts were initialized from the corresponding analyses at the synoptic times of 00, 06, 12, and 18 UTC and integrated for 12~h, with model state output every 30 minutes. For comparison, the ML forecasts were initialized at the same synoptic times using the most recent available satellite observations. Because MPAS-JEDI analyses assimilate observations valid up to 3~h after initialization, MPAS-A has a modest initialization advantage. However, this advantage has little impact on the comparison due to the performance characteristics of MPAS-A shown in the results.

Brightness temperatures were simulated from each 30-minute forecast output time using the Community Radiative Transfer Model (CRTM) version 2.4.1 in an all-sky configuration. Simulated fields were generated through the MPAS-JEDI HofX application, ensuring that the same observation operator, spatial interpolation, and quality control used during assimilation were applied for verification. Forecasts for all models were verified against GOES-19 ABI channel 13 observations at the matched valid times.

\subsection{ML models}

\begin{figure}[H]
  \noindent\includegraphics[width=\textwidth]{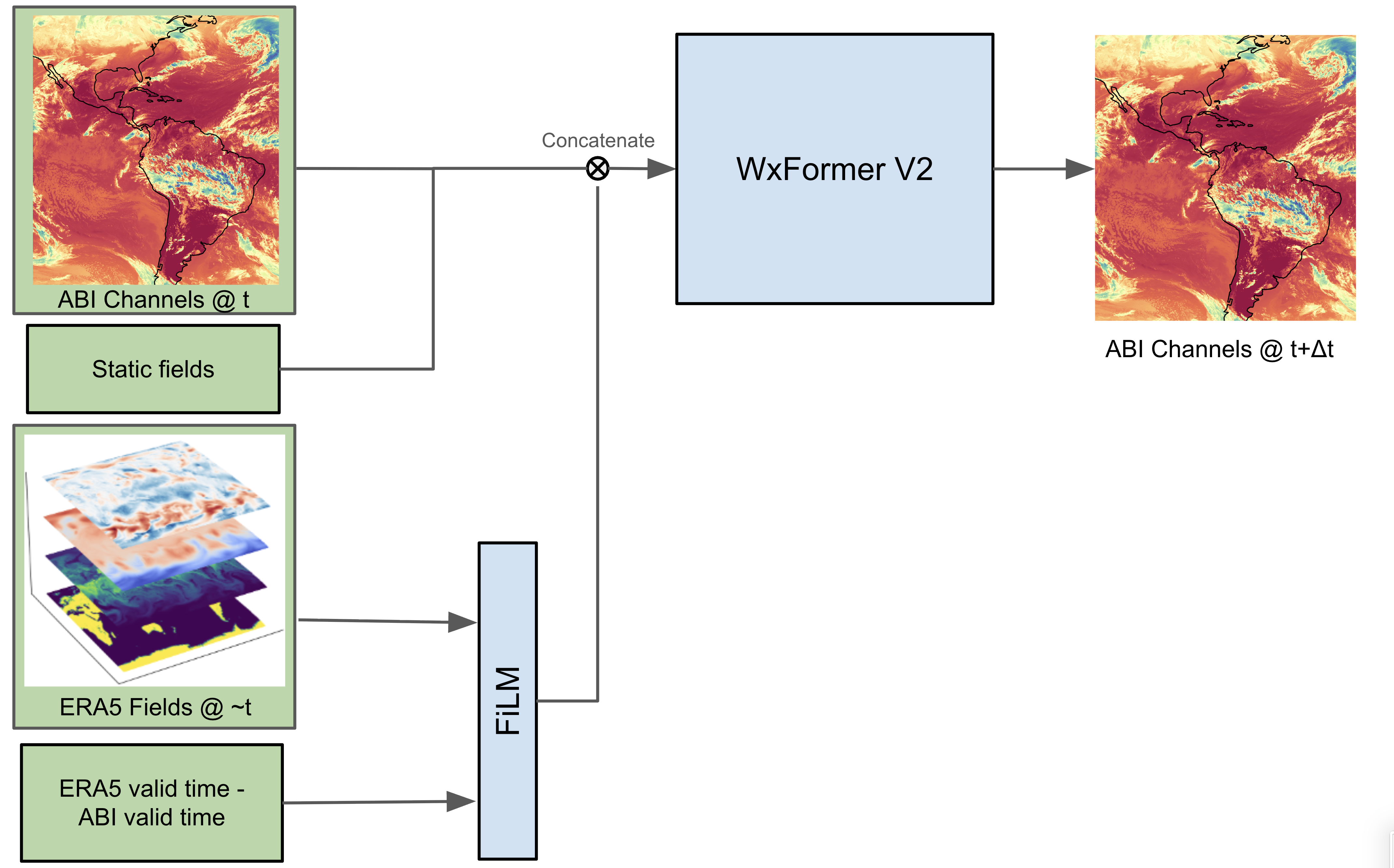}\\
  \caption{Illustration of the overarching model architecture and prediction domain. The learnable FiLM layer conditions the ERA5 fields according to the difference between the prediction time and ERA5 valid time. The results are concatenated with the ABI tensor before all being passed into the WxFormer V2 vision transformer backbone. Only ABI channels are predicted.}\label{fig:architecture}
\end{figure}

The DOP+ and DOP models are based on the WxFormer architecture \citep{schreck_community_2025}, a cross-scale transformer-based model for weather prediction, which we modify by replacing the final upsampling layers with super-resolution pixel shuffle layers. This modification enables high-resolution features to be recovered in the output without the checkerboard artifacts associated with transposed convolution upsampling \citep{shi_real-time_2016}. The full model architecture is illustrated in the supplementary material.

The DOP+ model accepts three inputs: GOES-East ABI imagery at 10-minute temporal resolution, static fields, ERA5 reanalysis fields at hourly resolution, and a scalar timedelta encoding the difference between the ERA5 field valid time and the GOES valid time. ERA5 is incorporated as a guidance forcing and is not a prediction target. The ERA5 fields are conditioned on the timedelta via a Feature-wise Linear Modulation (FiLM) layer \citep{perez_film_2018}, which learns to affinely transform the ERA5 features as a function of the timedelta to inform the model of this timedelta. The FiLM-conditioned ERA5 tensor is then stacked channel-wise with the GOES tensor and passed through the transformer backbone.

The model only outputs the GOES channels and the loss function is computed exclusively on the GOES channels. This is essentially treating the ERA5 fields as a prescribed forcing input rather than a prognostic variable.

At inference time, the model is applied autoregressively: the predicted GOES state at time $t + \Delta t$ is fed back as input alongside the ERA5 fields that are closest to that time, and the process is repeated to generate multi-step forecasts. The timedelta input allows the model to adjust to the difference between the true input time and the ERA5 valid time.

The DOP model operates similarly except that it accepts only GOES imagery, static fields, and TSI field as inputs.

\subsection{Training method and computational benchmarks}

All models were trained with a single-step prediction. This approach minimizes forecast smoothing which happens due to multi-step training. This approach maximized the ability for the model to represent fast phenomena, at the expense of some stability. Of course, the model could be tuned for other objectives, but here we optimized for representing fast phenomena.

The training objective is the latitude-weighted mean squared error (MSE) between the predicted and target GOES brightness temperatures and reflectance fields. Each pixel's contribution to the loss is weighted by the cosine of its latitude to account for the decrease in grid cell area with increasing distance from the equator. Although MSE incentivizes smoothing, our short 10 minute timestep is the same timestep used in the WoFSCast model in which ``despite using an MSE loss, no substantial loss of information at smaller scales occurs out to 2~hr'' \citep{flora_wofscast_2025}. This short timestep, combined with coarser resolution of the ABI fields, allows us to predict realistic fields further than the 2~hr horizon.

While we trained multiple models to convergence, the models featured in the evaluation all used a 10-minute timestep. These models are labeled DOP+ and DOP unless otherwise specified. We assessed the sensitivity of forecast skill to the model timestep by training several other models, where each model had a timestep of 10, 20, or 30 minutes. Ablation studies were also conducted to assess the impact of model complexity/size. These studies are available in the supplementary material.

DOP+ was trained for 45 epochs over 165632 samples with an effective batch size of 128. Other models were trained with 50--70 epochs. We used the Adam optimizer with a cosine annealing with restarts schedule. The models were trained with distributed data parallel on 32 A100-40GB GPUs using automatic mixed precision.

Model inference takes ${\sim}$0.5 seconds per timestep, including I/O. For a model with a 10 minute timestep, this means a 12~h forecast takes ${\sim}$40 seconds.

\nocite{karlbauer_advancing_2024}

\clearpage
\acknowledgments
We would like to thank Craig Schwartz for helpful discussions on the Fractions Skill Score of our models.

This research was supported by the United States Air Force (grant no.\ NA21OAR4310383) and the NSF National Center for Atmospheric Research, which is a major facility sponsored by the U.S.\ National Science Foundation under Cooperative Agreement No.\ 1852977.

\section*{Data availability}
The MPAS-JEDI cycling experiment simulated brightness temperatures are available at \url{https://app.globus.org/file-manager?origin_id=22275241-02e7-425a-89d6-5686435fdd46&origin_path=%2F}. The GOES ABI data is available at the AWS GOES archive \url{https://registry.opendata.aws/noaa-goes/}. The ERA5 model-level re-analysis data for this study can be accessed through the NSF NCAR Research Data Archive at \url{https://doi.org/10.5065/XV5R5344}.

\section*{Code availability}
The data pre-processing, neural networks, simulation, and verification code used to train, test, and verify the models are archived at \url{https://doi.org/10.5281/zenodo.21960122}.

\section*{Author contributions}
DK conceptualized the study, processed and curated the data, developed the software, trained and evaluated models, produced the visualizations, acquired computing resources, and wrote the original draft with contributions from all co-authors. OB aided in model training, ablation, and qualitative model selection; processed and regridded the ERA5, TSI, and static fields. IHB and BJ ran the MPAS-JEDI cycling experiment and generated the simulated brightness temperatures. CS conceptualized the study, developed the methodology, supervised evaluation, and acquired funding. All authors reviewed and edited the manuscript.

\section*{Competing interests}
The authors declare no competing interests.

\bibliographystyle{ametsocV6}
\bibliography{references}


\clearpage

\thispagestyle{empty}
\begin{center}
  \vspace*{4\baselineskip}
  {\Large\bfseries Supplementary Information\par}
  \vspace{1.5\baselineskip}
  {\large GOES-East full-disk AI nowcasting of cloud evolution in observation space\par}
\end{center}
\clearpage

\setcounter{section}{0}
\setcounter{figure}{0}
\setcounter{table}{0}
\setcounter{equation}{0}
\renewcommand{\thefigure}{S\arabic{figure}}
\renewcommand{\thesection}{S\arabic{section}}

\begin{figure}[t]
  \noindent\includegraphics[width=\textwidth]{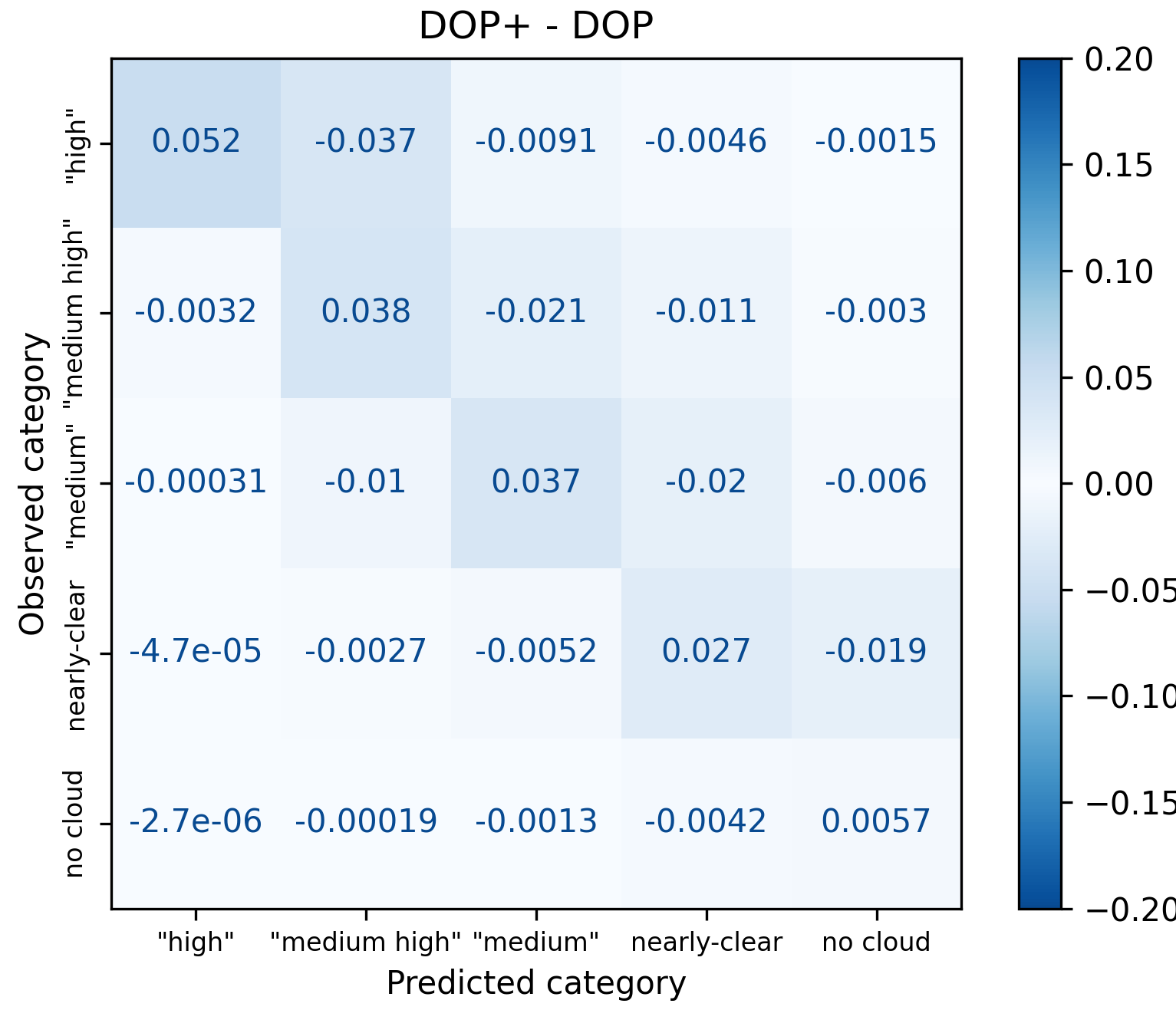}\\
  \caption{Difference of DOP+ and DOP row-normalized confusion matrices at 1~h lead time. Blue indicates an improvement (positive on the diagonal, negative on off-diagonal).}\label{fig:confusion_diff}
\end{figure}

\section{Ablation experiments}

\subsection{Timestep ablation}

\begin{figure}[t]
  \noindent\includegraphics[width=\textwidth]{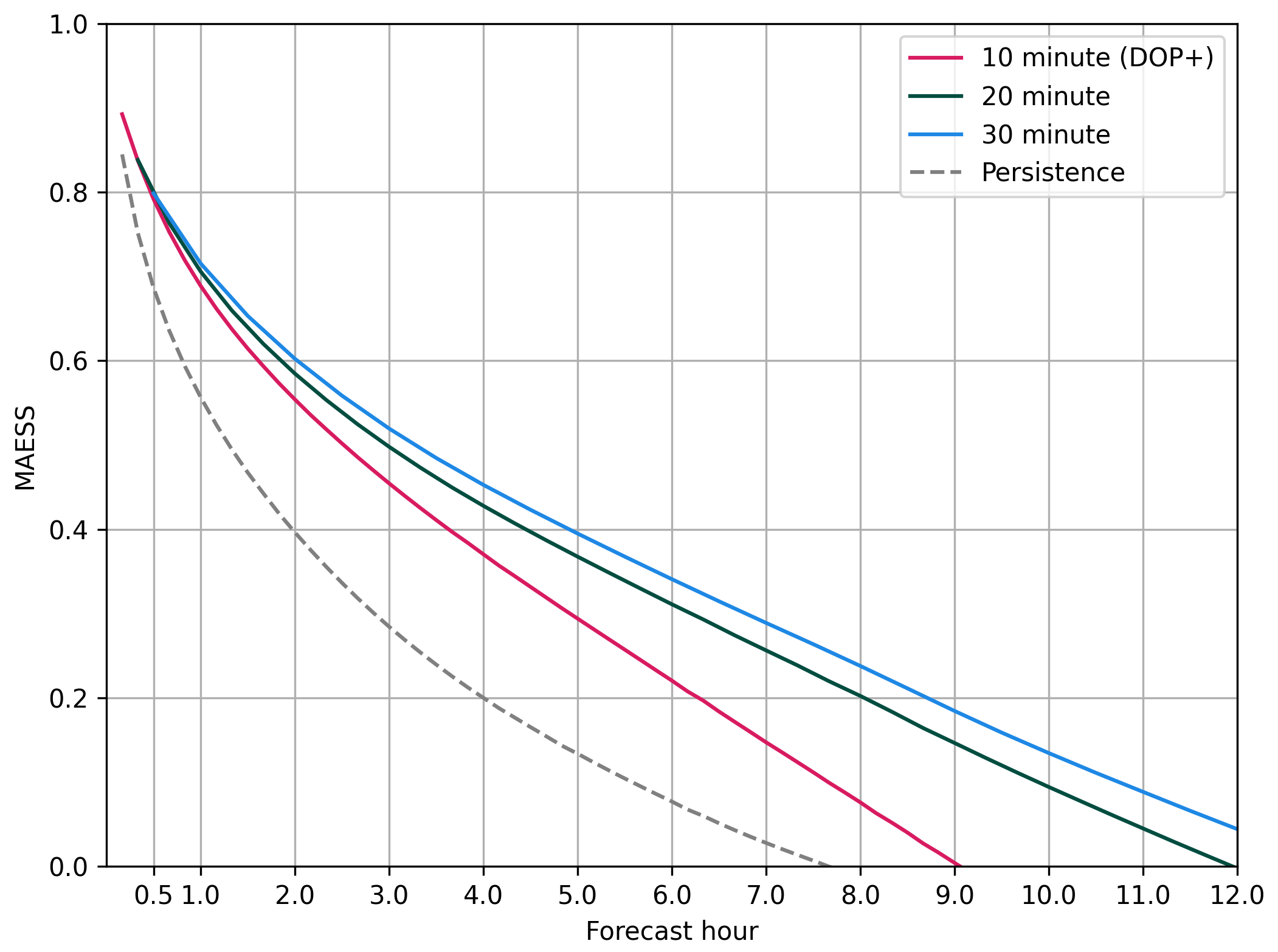}\\
  \caption{MAESS as a function of lead time for ABI band 13 for ML models with different timesteps.}\label{fig:maess_timestep}
\end{figure}

\begin{figure}[t]
  \noindent\includegraphics[width=\textwidth]{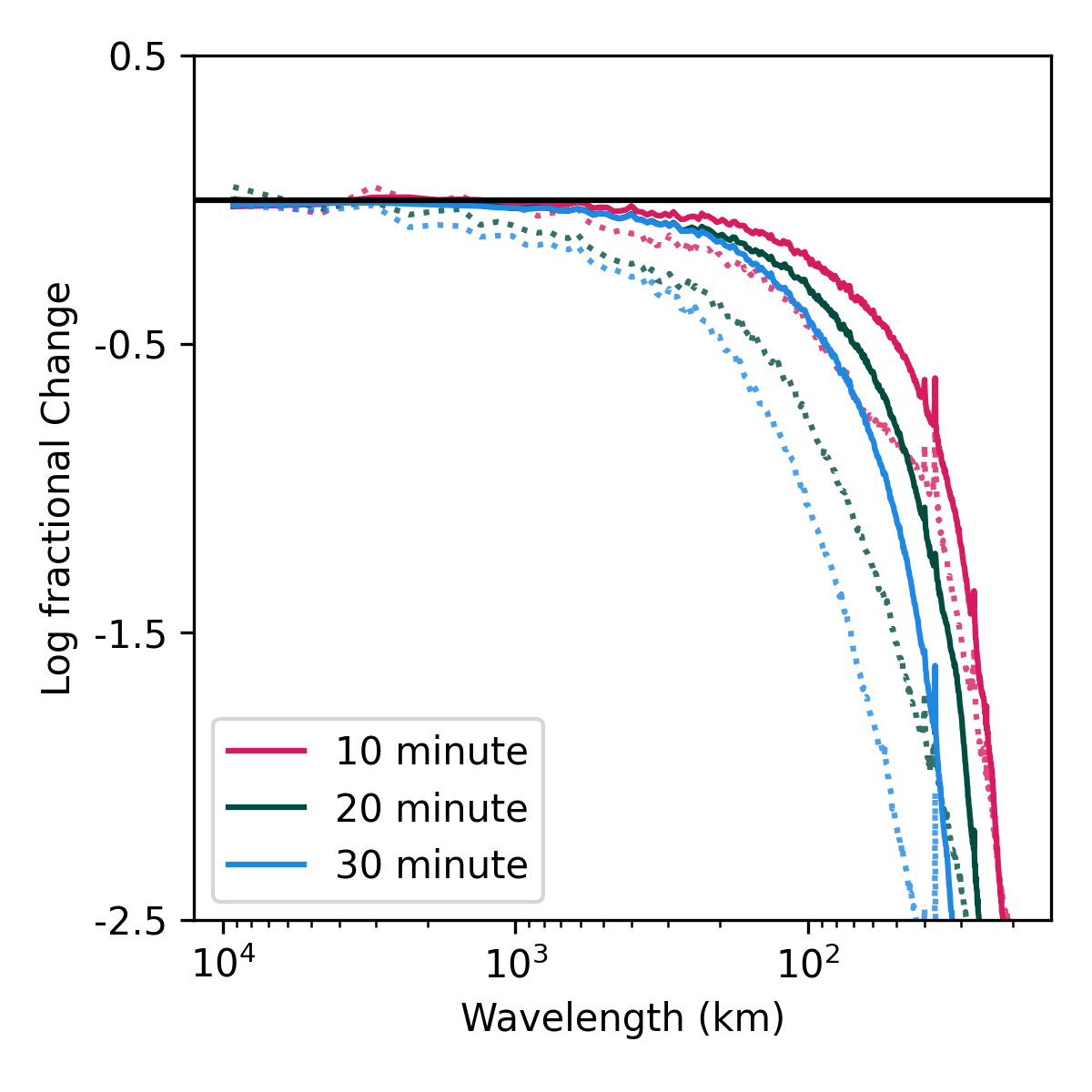}\\
  \caption{Log ratio of power spectral density (PSD) between forecast and reference (truth) fields for Band 13, as a function of spatial scale. Models with different timesteps are shown, keeping the architecture/model size fixed.}\label{fig:psd_timestep}
\end{figure}

\begin{figure}[p]
  \noindent\includegraphics[width=\textwidth]{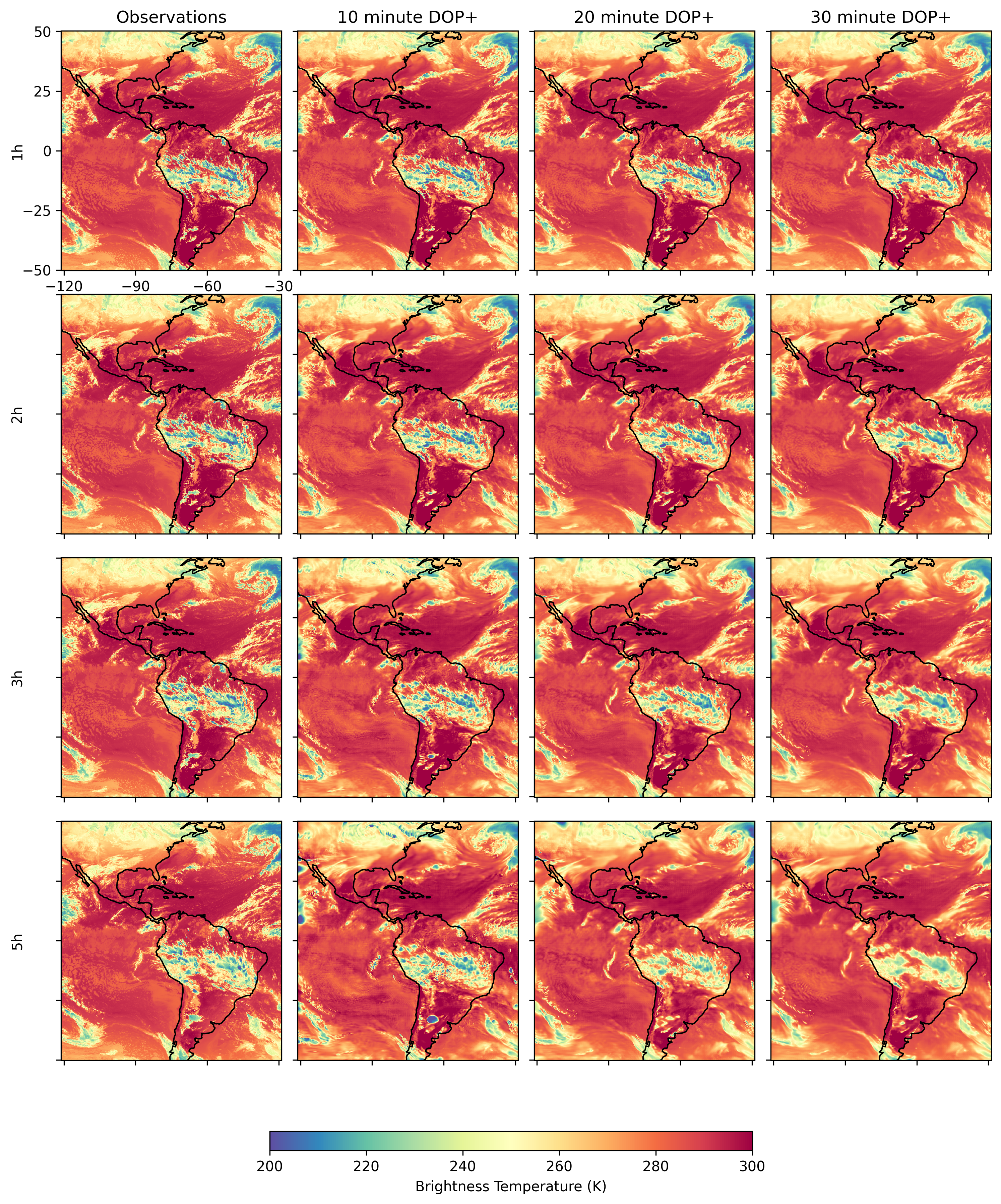}\\
  \caption{1, 2, 3, and 5 hour forecasts initialized at 1800 UTC December 16, 2022. Ablations with respect to the timestep are shown, where the model architecture and size were identical to DOP+.}\label{fig:ablation_timestep}
\end{figure}

We conducted ablation experiments to determine the effect of the prediction timestep and model capacity/complexity. Using the same model, three timesteps were evaluated, 10, 20, and 30 minutes (Figures \ref{fig:maess_timestep}, \ref{fig:psd_timestep}, \ref{fig:ablation_timestep}). We see a notable degradation in forecast sharpness as the timestep increases. This is consistent with MSE being minimized by smoothing where there is more uncertainty in the predicted fields. By subjective evaluation the 30 minute model has significantly less colder brightness temperatures than the 10 minute model. Thus the forecasted fields by the 30 minute model have very little realism in convective features.

\subsection{Model complexity ablation}

\begin{figure}[t]
  \noindent\includegraphics[width=\textwidth]{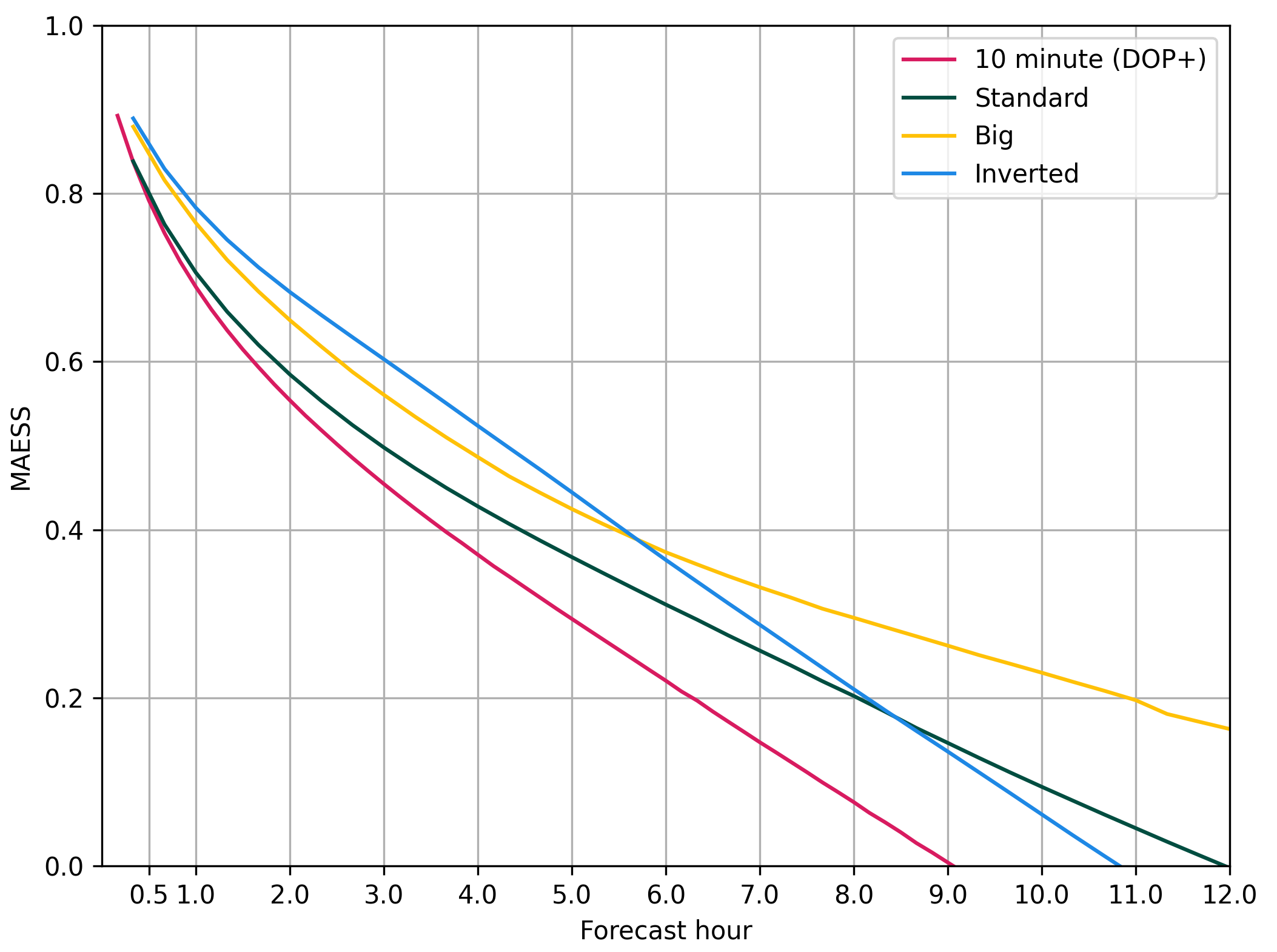}\\
  \caption{MAESS as a function of lead time for ABI band 13 for ML models of different complexities.}\label{fig:maess_model_size}
\end{figure}

\begin{figure}[t]
  \noindent\includegraphics[width=\textwidth]{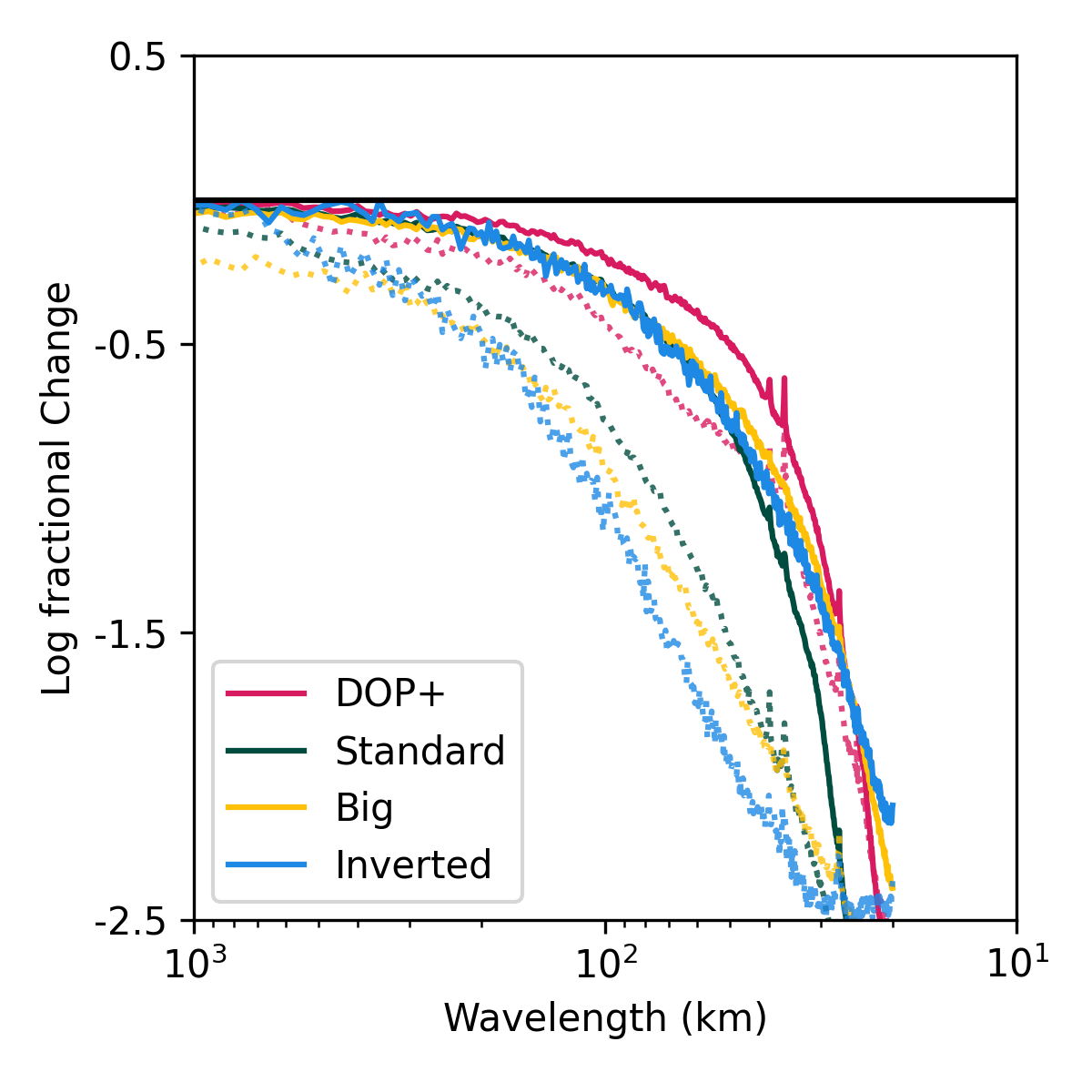}\\
  \caption{Log ratio of power spectral density (PSD) between forecast and reference (truth) fields for Band 13, as a function of spatial scale. Models of different complexities are shown.}\label{fig:psd_model_size}
\end{figure}

\begin{figure}[p]
  \noindent\includegraphics[width=\textwidth]{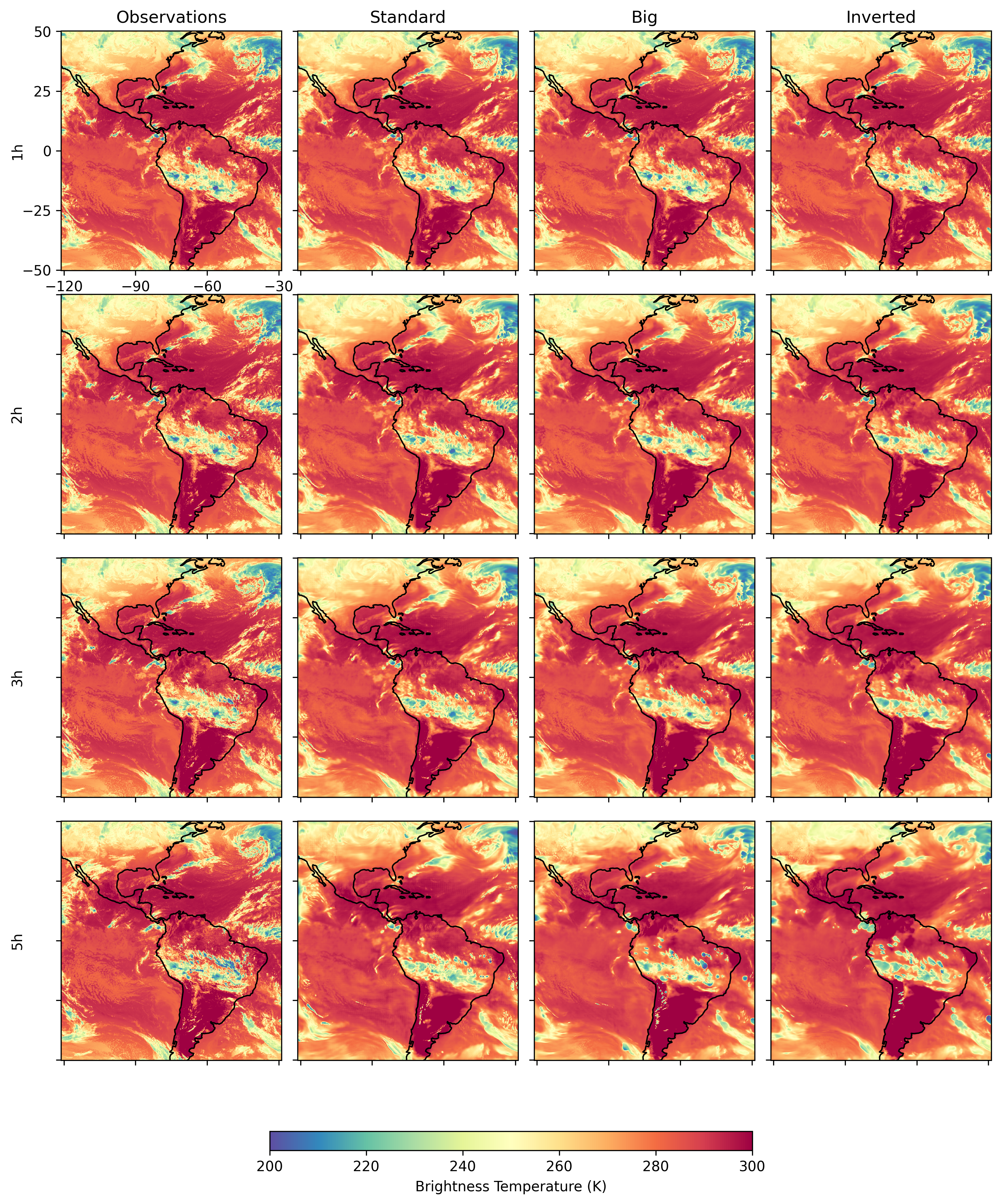}\\
  \caption{1, 2, 3, and 5 hour forecasts initialized at 1200 UTC December 16, 2022. Ablations with respect to the model complexity are shown. The prediction timestep was fixed at 20 minutes.}\label{fig:ablation_complexity}
\end{figure}

Using a 20 minute timestep, two more model configurations were chosen to assess the impact of model architecture and complexity (Figures \ref{fig:maess_model_size}, \ref{fig:psd_model_size}, \ref{fig:ablation_complexity}). The ``Big'' model uniformly increased the size of the latent dimension in the WxFormer pyramid architecture. The sequence of latent sizes is: [256, 512, 1024, 2048]. The ``Inverted'' model inverted the pyramid architecture, where the largest latent dimension is at the finest spatial scale, corresponding to a latent dimension sequence of [512, 256, 128, 64]. This inverted architecture was used in \citet{karlbauer_advancing_2024} due to the reasoning that fine grained weather features need more model capacity to encode them, while large scale features need less. Strictly speaking, the parameter count of the inverted model may be comparable to our standard configuration. However, the total data volume of the latent representations is much larger. Thus the model has an effectively more complex representation of the data.

\begin{figure}[p]
 \noindent\includegraphics[width=\textwidth]{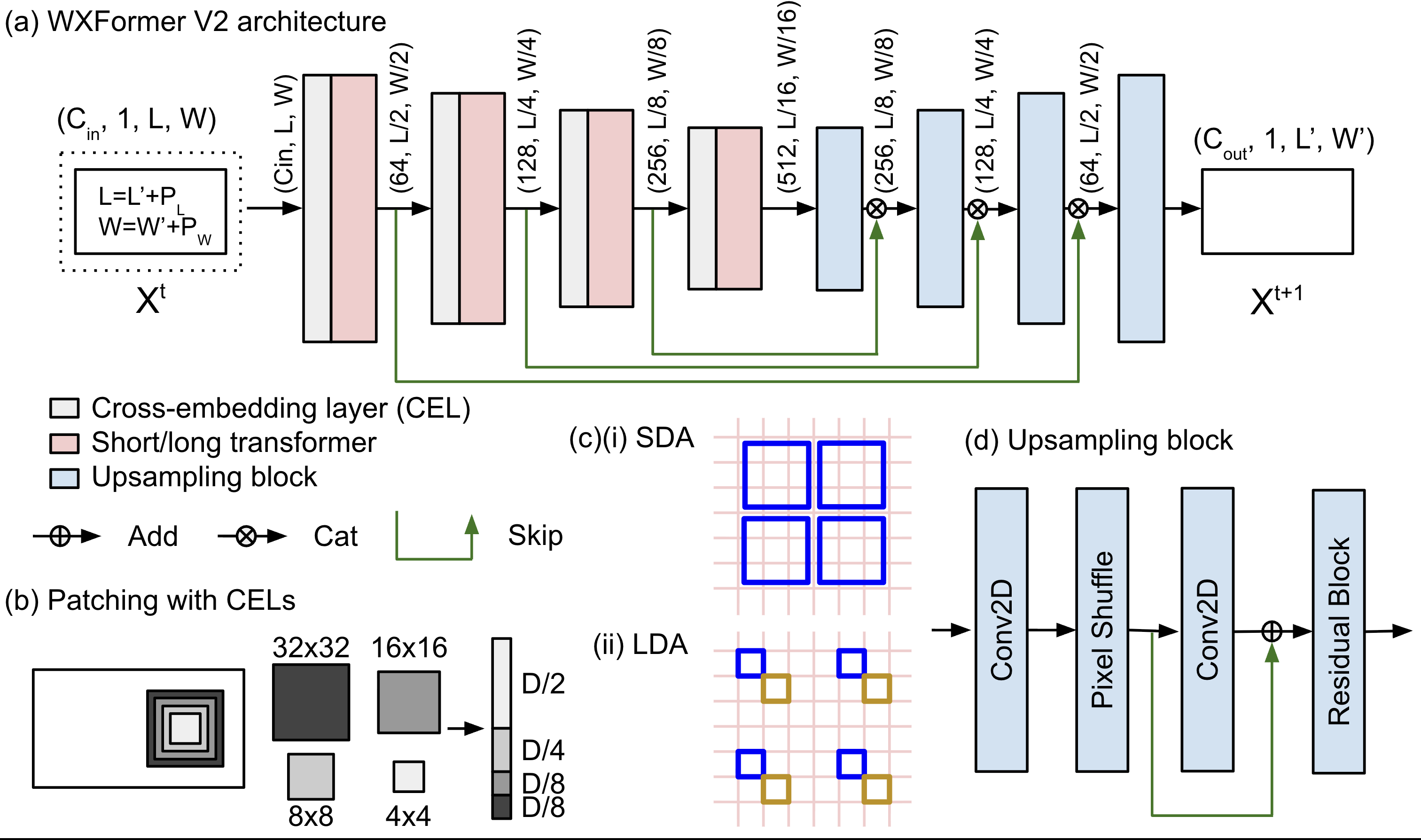}\\
 \caption{Overview of the WXFormer V2 architecture. (a) Encoder--decoder structure: encoding stages use a CrossFormer backbone, and decoding stages progressively restore spatial resolution, with skip connections linking corresponding levels. (b) The cross-scale embedding layer (CEL) extracts multi-scale features using four convolutional kernels of different sizes. (c) Long--short distance attention (LSDA), comprising (i) short-distance attention (SDA) for local interactions and (ii) long-distance attention (LDA) for global dependencies. (d) Decoder upsampling blocks, which use pixel shuffle to increase feature-map resolution, replacing the convolutional layer of V1.}\label{fig:wxformer_arch}
\end{figure}

\begin{figure}[t]
  \noindent\includegraphics[width=\textwidth]{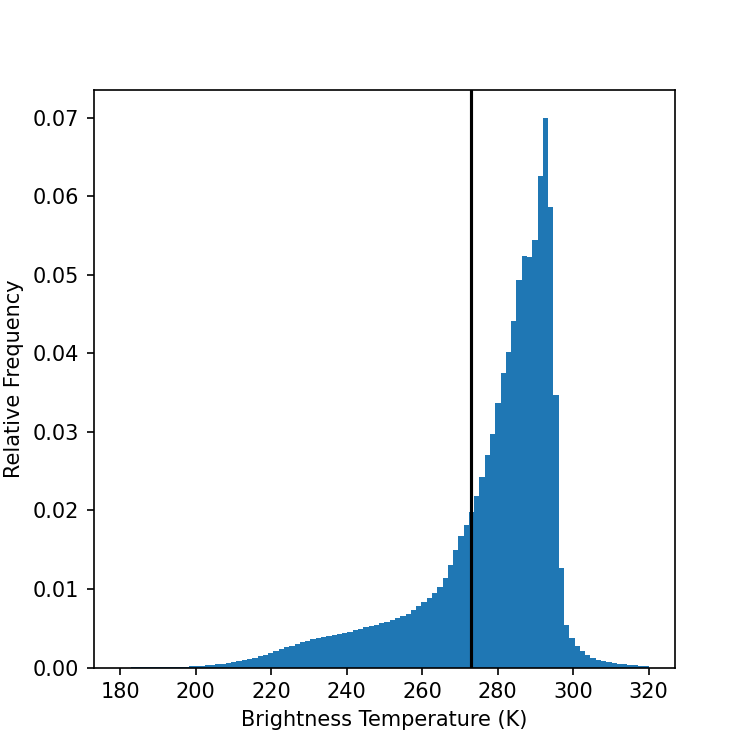}\\
  \caption{Histogram showing the distribution of brightness temperatures of band 13 over the evaluation period.}\label{fig:bt_hist}
\end{figure}

\end{document}